\documentclass[11pt]{article}

\usepackage[a4paper,margin=1in]{geometry}
\usepackage{setspace}
\usepackage[T1]{fontenc}
\usepackage[utf8]{inputenc}
\usepackage{lmodern}

\usepackage{amsmath}
\usepackage{amssymb}
\usepackage{amsfonts}
\usepackage{bm}

\usepackage{graphicx}
\usepackage{caption}
\usepackage{subcaption}
\usepackage{booktabs}
\usepackage{multirow}
\usepackage{array}
\usepackage{tabularx}
\usepackage{makecell}

\usepackage[round,authoryear]{natbib}
\usepackage[colorlinks=true,
            linkcolor=blue,
            citecolor=blue,
            urlcolor=blue]{hyperref}

\usepackage[switch]{lineno}

\usepackage{xcolor}
\usepackage{enumitem}
\usepackage{siunitx}

\title{\bfseries Breaking-induced energy dissipation of surface gravity waves at varying scales and co-flowing wind stresses}

\author{
\begin{minipage}{0.9\textwidth}
\centering
Rui Cao$^{1,2,3}$, Enrique M. Padilla$^{4}$, Xu Chen$^{1,2}$ and Adrian H. Callaghan$^{3,*}$\\[0.5em]
\small $^{1}$State Key Laboratory of Physical Oceanography, Ocean University of China, Qingdao 266100, China\\
\small $^{2}$College of Oceanic and Atmospheric Sciences, Ocean University of China, Qingdao 266100, China\\
\small $^{3}$Department of Civil and Environmental Engineering, Imperial College London, London SW7 2AZ, UK\\
\small $^{4}$Maritime Engineering Laboratory (LIM), Polytechnic University of Catalonia, Barcelona 08034, Spain\\[0.5em]
\small $^{*}$Corresponding author: \href{mailto:a.callaghan@imperial.ac.uk}{a.callaghan@imperial.ac.uk} \\
Contributing authors: \href{mailto:r.cao@ouc.edu.cn}{r.cao@ouc.edu.cn}, \href{mailto:enrique.padilla@upc.edu}{enrique.padilla@upc.edu}, \href{mailto:chenxu001@ouc.edu.cn}{chenxu001@ouc.edu.cn}
\end{minipage}
}

\date{13 November 2026}

\begin{document}
\maketitle

\begin{abstract}
Breaking-induced energy dissipation is studied for individual unsteady breaking waves using laboratory measurements of unidirectional surface gravity wave groups across a range of underlying wave scales and wind stresses. A refined framework to estimate the breaking-induced energy dissipation $\Delta E_{br}$ is first proposed which accounts for background dissipation arising from non-breaking processes. Using this framework, we show that variations in wave scale primarily influence breaking energetics, such as the fractional energy dissipation $\Delta E_{br}/E_0$ and the associated dissipation rate $\epsilon_b$, by modifying the breaking onset threshold. We also show that the presence of co-flowing wind systematically reduces both $\Delta E_{br}/E_0$ and $\epsilon_b$ compared to unforced conditions without wind, as wind-forced waves break earlier and have a reduced degree of crest forward-leaning. Exploiting the ability of the crest-front steepness at incipient breaking $\mathcal{S}_{\text{front}}(t_b)$ to characterise both the breaking onset threshold and the local crest geometry, an appropriate scaling for $\epsilon_b$ is subsequently formulated based on this local measure. This leads to a relation for fractional energy dissipation of the form $\Delta E_{br}/E_0 \propto \beta^{*}\,\mathcal{S}_b\,(\tau_b/T_b)$, where $\beta^{*}$, $\mathcal{S}_b$ and $\tau_b/T_b$ represent the degree of crest forward leaning, the local steepness, and the non-dimensional breaking duration. This scaling explicitly highlights that crest asymmetry and breaking duration play an important role in setting the breaking wave energy dissipation. Finally, we consider the implications for the breaking strength parameter $b$ by first assessing existing steepness-based scaling laws, and subsequently relating $b$ to $\mathcal{S}_{\text{front}}(t_b)$ which yields an approximately linear dependence above a breaking-onset threshold.
\end{abstract}

\section{\label{sec: intro}Introduction}

The surface wave energy balance comprises three primary processes: energy input from the wind, non-linear energy re-distribution across different underlying wave components and energy dissipation \citep{Hasselmann1962}. Among these, energy dissipation is to a large extent controlled by wave breaking, whose underlying physics remains incompletely understood \citep{Drazen2008,Herman2026}. Consequently, wave breaking acts as the primary mechanism limiting the growth of surface waves and regulating their evolution under continuous wind forcing \citep{Hasselmann1962,Melville1996,Perlin2013,Callaghan2018, Hogan2025}. In a broader context, wave breaking plays a direct role in air--sea exchange processes and provides a key physical constraint for spectral wave modelling \citep{Deike2022, Callaghan2025whitecap}. Accurately characterising both the severity and the rate of energy dissipation associated with individual breaking events therefore remains an active area of research.

\subsection{\label{subsec: Background}Background}
Owing to the considerable technical challenges involved in directly resolving the detailed energetics of individual breaking events in the open ocean, with only very limited observations reported to date \citep[e.g.][]{Callaghan2025b}, much of our present knowledge of breaking-induced energy dissipation (rates) has been derived from controlled laboratory experiments \citep[e.g.][]{Duncan1981, Ramberg1987, Rapp1990, Kway1998, Drazen2008, Tian2010, Tian2012, Callaghan2013, Callaghan2016, Deane2016, Craciunescu2020, Sinnis2021, Cao2023, Cao2026Representative} and numerical simulations \citep[e.g.][]{Iafrati2011, Derakhti2014, Derakhti2016, Deike2016, Vita2018, Mostert2022, Liu2023, Scapin2026}. Commonly, energy dissipation has been expressed by means of a non-dimensional form, defined as the ratio of the total energy lost during breaking, $\Delta E_{br}$, to the initial energy contained within the wave group, $E_{0}$. Across a wide range of breaking conditions, this fractional energy loss has been reported to lie between $\Delta E_{br}/E_{0} \sim \mathcal{O}(1\%)$ and $\mathcal{O}(10\%)$ \citep{Rapp1990,Kway1998,Zhang2005,Tian2010,Allis2013}.

In a similar spirit, energy dissipation rates associated with wave breaking have often been non-dimensionalised following the framework introduced by \citet{Duncan1981}, in which the dissipation rate is characterised by the breaking strength parameter,

\begin{equation} \label{eq: b}
b = (g\rho_w^{-1}) \epsilon_b c^{-5}.
\end{equation}
Here, $g$ is the gravitational acceleration, $\rho_w$ is the water density, $\epsilon_b$ is the rate of energy dissipation per unit crest length (spanwise-averaged) due to wave breaking, and $c$ represents a measure of the breaker's phase speed. Considerable effort has been devoted to constraining $b$, motivated by the prospect that, once its dependence on any more readily measurable quantity is established \citep[e.g.][]{Drazen2008}, measurements of $c$ alone may be sufficient to infer the magnitude of $\epsilon_b$ using \eqref{eq: b} across a range of breaking scales.

A combination of both physical arguments and experimental measurements have consistently shown that wave steepness, $\mathcal{S}$, is the primary geometric variable controlling variations in the breaking-induced, dimensionless energy dissipation ($\Delta E_{br}/E_{0}$) as well as the breaking strength parameter ($b$) \citep{Rapp1990,Drazen2008,Romero2012,Perlin2013,Deike2022}. While wave steepness has been defined in different ways across studies, it is generally constructed from a combination of some characteristic wave amplitude, $a$, and some characteristic wavenumber, $k$, such that $\mathcal{S} \sim a k$. Notwithstanding differences in the precise definition of wave steepness, there is broad agreement that, for individual breaking events, $\Delta E_{br}/E_{0}$ increases monotonically with $\mathcal{S}$ at low steepness and may approach an upper bound as breaking becomes sufficiently energetic or involves multiple breaking processes. The breaking strength parameter, $b$, has likewise been reported to be constrained by a power-law dependence on wave steepness, 

\begin{equation} \label{eq: power-law}
b \propto \mathcal{S}^{\Xi},
\end{equation}
where the exponent $\Xi=5/2$ is obtained when inertial scaling arguments for turbulent dissipation are adopted \citep{Drazen2008}. This scaling, or variations thereof, have been followed in related studies \citep{Romero2012,Grare2013,Deike2016,Vita2018,Craciunescu2020,Sinnis2021,Mostert2022,Scapin2026}. In these studies, wave steepness was defined explicitly as: 

\begin{equation} \label{eq: Sn}
\mathcal{S} \sim \mathcal{S}_n = \sum_{i=1}^{N} a_i k_i, 
\end{equation}
which represents the maximum linear target steepness of a propagating wave group composed of $N$ underlying components. Reported values of $\Xi$ in equation~\eqref{eq: power-law} are, however, sensitive to the specific definition of $\mathcal{S}$ employed, with smaller exponents in the range $\Xi \in [1,\,2.5]$ obtained in other formulations \citep{Tian2010,Derakhti2016,Cao2023}, highlighting the challenge when comparing results from different studies that employ different underlying measures of wave steepness.

\subsection{\label{subsec: Objectives} Problem formulation and objectives}
Although breaking-induced energy loss and dissipation rates are commonly non-dimensionalised and scaled using wave steepness owing to its close link with wave non-linearity and breaker intensity, available evidence indicates that an intrinsic scale dependence remains. 

For example, for a given value of $\mathcal{S}_n$, breaking waves of different scales have been observed to exhibit systematically different levels of fractional energy dissipation, $\Delta E_{br}/E_{0}$, as evident in figure~3(a) of \citet{Perlin2013}. In addition, laboratory experiments on two-dimensional, unsteady, dispersively focused breaking waves by \citet{Drazen2008} showed that larger-scale breakers, characterised by longer wavelengths (or smaller frequencies), generally dissipate a greater fraction of their energy than smaller-scale breakers, despite being generated at comparable target steepness (see their figure~8). In that study, wave scale was characterised using the central frequency, $f_c$, of the constant-steepness spectra employed. Evidence of a similar scale dependence has also been reflected in previous $b$--$\mathcal{S}_n$ relationships. For example, \citet{Derakhti2016} numerically reproduced selected breaking wave groups from \citet{Drazen2008} at $f_c = 0.88\,\mathrm{Hz}$ and from \citet{Tian2012} at $f_c = 1.7\,\mathrm{Hz}$, and showed in their figure~14($b$) that values of $b$ associated with larger-scale cases are systematically larger at comparable $\mathcal{S}_n$ than those associated with smaller-scale cases. It is important to note that, in the experiments of \citet{Drazen2008} and the corresponding numerical reproductions considered by \citet{Derakhti2016}, variations in $f_c$ are naturally accompanied by changes in spectral bandwidth, such that the effects of wave scale and bandwidth may not be fully independent. One aim of the present work is therefore to isolate the role of wave scale and to examine its influence on breaking-induced energy dissipation (and its rates) using a relatively wide range of underlying wave conditions.

In addition to wave scale, previous studies have highlighted the influence of a range of wave and spectral parameters that can modify the relationships between $\Delta E_{br}/E_{0}$, $b$, and  $\mathcal{S}$. These include wave directionality \citep{Wu2002,Allis2013}, the mechanism of wave group formation \citep[e.g. dispersive focusing versus modulational instability,][]{Banner2007,Allis2013,Derakhti2016}, spectral bandwidth \citep{Sinnis2021,Cao2023}, water depth \citep{Liu2023}, and the presence of background currents \citep{Wu2004}. It remains unclear, however, how direct wind forcing alters these relationships, despite the fact that the vast majority of breaking events in the ocean occur under wind-forced conditions. To date, only a limited number of studies have explicitly examined the influence of wind on breaking-induced energy dissipation, and the findings reported are mixed. For example, the numerical study of \citet{Iafrati2019} found no significant difference in fractional energy dissipation between wind-forced and non-wind-forced cases. Very recently, the DNS results of \citet{Scapin2026} showed that the $b \propto \mathcal{S}_n^{5/2}$ scaling remains valid for wind-forced breakers. In contrast, the combined experimental and numerical results of \citet{Galchenko2012} suggested that direct wind forcing can reduce the energy loss of individual breaking events while increasing the likelihood of breaking.

In our previous work \citep{Cao2025prep}, among others \citep[e.g.][]{Saket2017,Knobler2022}, the manner in which direct wind stress modifies the wave shape at incipient breaking has been examined. In particular, wind forcing was found to reduce the forward leaning and crest-front steepness of the limiting crest. The implications of these geometric changes for breaking-induced energy loss, the associated dissipation rate and $b$, however, have not been quantified. Here, we aim to systematically examine how direct wind stress influences these energetic quantities and to interpret the observed effects in the context of wind-induced modifications to the local crest geometry.


With the above aims in mind, we use extensive laboratory measurements to address outstanding questions regarding the roles of wave scale and direct (co-flowing) wind stress in controlling the energy dissipation associated with breaking surface gravity waves. Details of the experimental datasets are provided in \S\ref{sec: experiment}. In \S\ref{sec: method}, we introduce a refined framework for quantifying the breaking-induced energy loss, $\Delta E_{br}$, in laboratory conditions, alongside a brief reassessment of methods employed in previous studies. This framework is designed to isolate $\Delta E_{br}$ more accurately by accounting for non-breaking energy losses arising during wave group propagation in the flume. The experimental results and their interpretation are presented in \S\ref{sec: Result}, where we relate our newly quantified breaking energetics to the local crest geometry identified in our earlier work \citep{Cao2025prep}. We establish a relationship for the energy dissipation (rate) across wave scales and wind-forcing conditions, and derive a scaling for fractional energy loss that highlights the combined roles of local non-linearity, crest leaning and non-dimensional breaking duration. We further identify an approximately linear relationship between $b$ and a form of the local crest steepness. A summary of the key findings and implications is given in \S\ref{sec: conclusion}.


\section{\label{sec: experiment} Experimental datasets: \textit{SIREN}, \textit{BUBER} and \textit{EURUS}}
The laboratory measurements of surface breaking waves analysed in the present study were obtained from three experimental campaigns, formally referred to as \textit{SIREN}, \textit{BUBER}, and \textit{EURUS}. All three campaigns were conducted in the Hydrodynamics Laboratory of the Department of Civil and Environmental Engineering at Imperial College London between 2020 and 2021. A detailed description of the experimental facilities, wave generation procedures, wave conditions, and measurement systems can be found in \S2 of \citet{Cao2025prep}. Here, we provide only a brief overview of the measurements and wave group configurations directly relevant to the present work and refer the reader to that study for further details.


\subsection{\label{subsec: generation} Generation of wave groups}
Experiments in all three campaigns were carried out in a glass-walled, wind--wave flume with dimensions of 27\,m in length, 0.30\,m in width, and a constant water depth of $d=0.70$\,m, equipped with flap-type paddles at either end of the flume. Individual breaking waves were generated within dispersively-focused, unsteady wave groups following the technique of \citet{Rapp1990}. This technique allows energetic or breaking events to be produced at a prescribed location and time by appropriately tuning the phases of the constituent Fourier components. In our experiments, breaking was targeted at a streamwise location of $x\approx9.5$\,m from the wavemaker.

Wave groups were generated using JONSWAP-type NewWave variance-density spectra, and their properties were characterised by three spectral parameters: (i) the linear amplitude sum of all Fourier components, $A=\sum_{i=1}^{N} a_i$, which controls the degree of wave-group nonlinearity; (ii) the peak enhancement factor, $\gamma$, which determines the concentration of spectral energy around the peak frequency $f_p$ and thus controls the spectral bandwidth; and (iii) the peak wave period, $T_p = 1/f_p=2 \pi/ \omega_p$, which sets the characteristic spectral scale. The corresponding peak wavenumber, $k_p$, and wavelength, $\lambda_p=2\pi/k_p$, are obtained from the finite-depth linear dispersion relation $\omega_p^2=gk_p\tanh(k_pd)$. Values of $T_p$ were varied to modify the characteristic wave scale, with larger values of $T_p$ corresponding to smaller $k_p$ and longer $\lambda_p$.

We use JONSWAP-type spectra because they are more representative of realistic ocean wave conditions \citep{McAllister2024}, while allowing the wave scale to be adjusted through $T_p$ with a prescribed spectral bandwidth set by $\gamma$ and an approximately self-similar spectral shape \citep{Cao2025prep}.

\subsection{\label{subsec: wave group details} Wave group details}
Consistent with the objectives outlined in \S\ref{subsec: Objectives}, the \textit{SIREN} campaign was designed to investigate the role of wave scale, whereas the \textit{BUBER} and \textit{EURUS} campaigns were used to examine the effects of direct wind forcing. The specific wave-group parameters employed in each campaign, including $A$, $\gamma$, and $T_p$, are summarised in table~1 of \citet{Cao2025prep}. Briefly, wave groups in all three campaigns had a fixed repeat period of 64\,s, resulting in a uniform frequency resolution of $\delta f = 1/64$\,Hz. The discrete frequency components spanned the range from a fixed lower bound of $f = 0.406$\,Hz to an upper bound of $f = 3 f_p$, giving a total number of spectral components
$N=1+(3f_p-0.406)/\delta f$.

In the \textit{SIREN} campaign, we varied the peak wave period over the range $T_p \in [1.0,\,1.6]$\,s in increments of 0.1\,s for cases with $\gamma = 2$, in order to cover a broad range of wave scales while keeping the spectral shape self-similar. In particular, this range of $T_p$ corresponds to peak wavelengths increasing from $\lambda_p=1.55$ to $3.43$\,m. Additional cases with $T_p = [1.1, 1.3, 1.5]$\,s were considered for $\gamma = 3$. Although spectral bandwidth can significantly affect the breaking wave geometry and energetics \citep{Cao2023,Mcallister2023}, its influence is not the focus here and so we considered only $\gamma=2$ and 3, corresponding to $\varepsilon_1=0.271$ and 0.254, respectively, where $\varepsilon_1=\sqrt{(m_1m_{-1})m_0^{-2}-1}$ \citep{Saulnier2011}. Here, $m_n$ denotes the $n$th-order spectral moment of the surface elevation variance-density spectrum, $S_{\eta\eta}(f_i) = a_i^2 / (2 \delta f)$. For each set of $(T_p, \gamma)$, the linear amplitude sum $A$ was progressively increased from $A = 20$\,mm up to the largest value for which isolated breaking events could be produced (i.e. we did not consider multiple breaking waves within a single wave group), thereby covering a range of non-breaking and breaking wave groups.

For the other two campaigns (\textit{BUBER} and \textit{EURUS}), wave groups were generated either without wind or under continuous co-flowing wind forcing, with $T_p = 1.2$ and 1.3\,s, and $\gamma = 2$ and 3. The \textit{BUBER} dataset corresponds to the no-wind reference cases, whereas the \textit{EURUS} dataset includes two wind speed conditions with 10-m equivalent wind speeds of $\overline{U_{10}} \approx 3.0$ and 6.0\,m\,s$^{-1}$, respectively. $\overline{U_{10}}$ was obtained by fitting logarithmic profiles to the time-averaged turbulent wind speed measurements at a fetch of $x = 9.4$\,m using a TSI-8455-075-1 wind probe, and extrapolating the fits to the standard 10-m reference height assuming neutral atmospheric conditions (see \S2.3.2 in \cite{Cao2025prep} for details). The corresponding local wave ages were $C_p/\overline{U_{10}}\approx0.6$ and 0.3, through which we related the laboratory conditions to relevant wind-wave conditions in the field. In all campaigns, the breaking location was at $x \approx 9.5$\,m.

\subsection{\label{subsec: measurements} Measurements}

With regard to measurements relevant to the present study, the evolution (surface elevation) of propagating wave groups upstream and downstream of the breaking region was recorded using an array of drop-down type, resistive wave gauges, sampling at 128\,Hz. A total of 14 gauges were deployed in the \textit{SIREN} campaign, and 16 gauges in the \textit{BUBER} and \textit{EURUS} campaigns. No wave gauges were placed within the immediate breaking region. Instead, the breaking process itself, together with its upstream and downstream spatial extent, was captured using three side-view digital cameras. Image sequences were recorded at 52\,Hz for \textit{SIREN} and at 20\,Hz for \textit{BUBER} and \textit{EURUS}, covering a streamwise distance of approximately 4.5\,m, from $x \approx 8.0$ to 12.5\,m.

Image-processing tools developed in \citet{Cao2025} (CMG) and in \citet{Cao2025prep} (SDBW-I) were employed to extract the spatial profile of the free surface at incipient breaking. This measurement is essential for determining the local wave steepness, which will be defined in \S\ref{subsubsec: steepness definitions}. In addition, a hydrophone operating at a sampling frequency of 250\,kHz was installed beneath the breaking region to record acoustic emissions associated with bubbles. Following established approaches \citep{Deane2002,Drazen2008,Sinnis2021, Zou2022}, the acoustic records were processed to identify the duration $\tau_b$ over which breaking remained acoustically active. Details of the hydrophone signal processing procedure are provided in \citet{Cao2026Representative}. This timescale was subsequently used in conjunction with the breaking-induced energy loss, $\Delta E_{br}$, to infer the corresponding energy dissipation rate, $\epsilon_b$ (as detailed in \S\ref{subsec: tau_b}).

\section{\label{sec: method}Methods for quantifying breaking-induced energy dissipation in laboratory flumes}  

Approaches used to quantify the energy dissipated by individual breaking waves, $\Delta E_{br}$, in finite-width, unidirectional laboratory wave flumes have been varied across studies. In this section, we consider three representative methods, proposed by \citet{Drazen2008}, \citet{Tian2010} (building on earlier ideas by \citet{Banner2007}), and \citet{Cao2023}, hereafter referred to as D08, T10, and C23.

The D08 method employs a control volume framework to estimate the loss of wave energy associated with breaking, within which background dissipation (e.g. viscous losses due to flume sidewall friction) is implicitly accounted for. As discussed in \citet{Cao2023}, however, the treatment of background energy losses in this approach is approximate and may lead to systematic overestimations in the values $\Delta E_{br}$ computed. The T10 method adopts a different strategy, in which a dense array of wave gauges distributed along the flume is used to directly characterise the evolution of wave energy and associated background dissipation. However, it requires extensive instrumentation and may not always be practical to implement, depending on the experimental configuration. Building on the control volume framework of D08, C23 revisited the representation of background dissipation and introduced a physically motivated correction that improves the accuracy of $\Delta E_{br}$ calculations.

In the following, we comment on these existing approaches and their underlying assumptions, before introducing a refined framework based on C23 that is adopted in the present study.

\subsection{\label{subsec: energy flux} General framework for wave energetics and breaking-induced energy dissipation}

In quantifying the amount of energy dissipated by wave breaking in unidirectional wave groups, one first considers the local mechanical energy flux, $\mathcal{F}(x,z,t)$ (units: J$\;$m$^{-2\;}$s$^{-1}$), through vertical sections at different streamwise locations along the wave flume. This flux represents the rate at which mechanical energy is transported in the streamwise direction, including contributions from kinetic energy, gravitational potential energy and pressure work. Mathematically, it may be expressed as

\begin{equation} \label{eq:local flux}
\mathcal{F}(x,z,t)=\left(\frac{1}{2}\rho_{w}\mathbf{U}^{2}+\rho_{w}gz+\mathcal{P}\right)u_x,
\end{equation}
where $\mathbf{U}=\sqrt{u_x^{2}+u_z^{2}}$ is the magnitude of the fluid velocity, $z$ denotes the vertical coordinate measured upward from the still water level, and $\mathcal{P}$ is pressure. Hence, the two-dimensional total wave energy passing through the section at $x$ over a finite time interval $\Delta T$, denoted as $E(x)$ (units: J\,m$^{-1}$), can then be written as the time-integrated, depth-integrated energy flux through that section,

\begin{equation} \label{eq:fluid energy}
E(x)=\int^{\Delta T}\int_{-d}^{\eta}\mathcal{F}(x,z,t)\,\mathrm{d}z\,\mathrm{d}t
=\int^{\Delta T}\int_{-d}^{\eta}\left(\frac{1}{2}\rho_{w}\mathbf{U}^{2}+\rho_{w}gz+\mathcal{P}\right)u_x\,\mathrm{d}z\,\mathrm{d}t.
\end{equation}
Under the assumptions of linear wave theory, for which the averaged depth-integrated kinetic and potential energy densities are assumed to be equal, $E(x)$ may be estimated from point measurements of the surface elevation time series, $\eta(x,t)$, according to

\begin{equation} \label{eq:eta energy}
E(x)=\rho_{w}g\,\langle C_g\rangle(x)\int^{\Delta T}\eta^{2}(x,t)\,\mathrm{d}t,
\end{equation}
where $\langle C_g\rangle(x)$ represents a measure of the wave group velocity that is a function of space \citep{Derakhti2016,Xu2022,Cao2023}. This equipartition assumption is valid only when the local non-linearity of the wave group remains sufficiently weak, and therefore equation~\eqref{eq:eta energy} is not expected to be strictly applicable in the vicinity of focal or breaking locations, where non-linear effects can become large.

\begin{figure*}
    \centering
	\includegraphics[width=0.55\columnwidth]{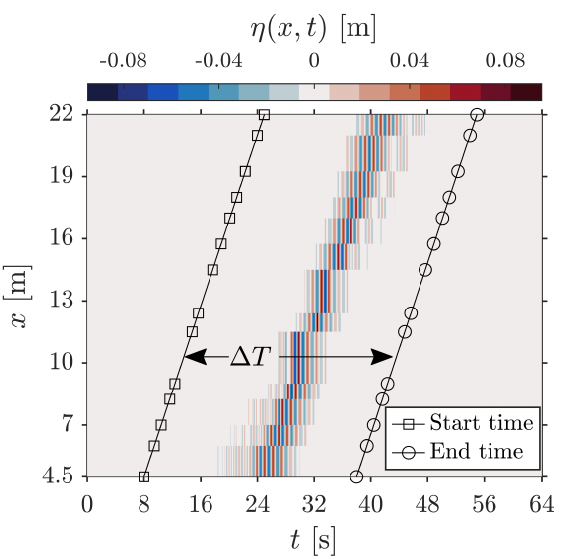}
	\caption{\label{fig: Delta T} Illustration of the definition of the integration time window, $\Delta T=30$\,s, used in equation~\eqref{eq:eta energy}. The example shown corresponds to a wave group from the \textit{SIREN} campaign with $\gamma=2$, $T_p=1.2$\,s, and $A=80$\,mm. The time domain plotted covers the full 64\,s repeat period, and the start and end times of the integration window are indicated, with squares and circles marking the streamwise locations of the 14 wave gauges used to measure the surface elevation. The integration window translates at a speed equal to the target wave group velocity, $C_{gp}=\partial\omega_p/\partial k_p=\omega_p\bigl(1+1.4k_p/\sinh(1.4k_p)\bigr)/(2k_p)$, such that the entire wave group is captured as it propagates through the measurement array, while any contributions outside the window are excluded.}
\end{figure*}

In the present study, $\langle C_g\rangle(x)$ in equation~\eqref{eq:eta energy} was evaluated locally as a spectral-composition-weighted group velocity, following the approach originally proposed by \citet{Drazen2008} (see their equation~(4.5)),

\begin{equation} \label{eq:Cgs}
\langle C_g\rangle(x)\sim C_{gs}(x)
=\frac{\displaystyle \sum_{i=1}^{N}C_{g,\, i}\,a_i^{2}\,\delta f}{\displaystyle \sum_{i=1}^{N}a_i^{2}\,\delta f},
\end{equation}
where $C_{g,\, i}$ represents the linear group velocity of the $i$th Fourier component. The integration duration $\Delta T$ in equation~\eqref{eq:eta energy} was chosen as a fixed time window of 30\,s (out of the full 64\,s repeat period), which was sufficiently long to capture the entire propagating wave group at a given gauge while minimising contamination from reflected wave energy (see figure~\ref{fig: Delta T}), although the independently estimated reflection coefficient was already small and below 3\% \citep{Cao2024}.

The total energy loss over the breaking process, $\Delta E$, is obtained by considering a control volume between the inlet $x_I$ and outlet $x_O$:

\begin{equation} \label{eq:delta E 1}
\Delta E = E(x_I)-E(x_O).
\end{equation}
This total energy loss includes both breaking-induced dissipation ($\Delta E_{br}$) and background dissipation ($\Delta E_{fr}$) as previously mentioned, giving

\begin{equation} \label{eq:delta E 2}
\Delta E = \Delta E_{br}+\Delta E_{fr}.
\end{equation}
Combining equations~\eqref{eq:delta E 1} and \eqref{eq:delta E 2}, the breaking-induced energy dissipation is therefore

\begin{equation} \label{eq:delta E br}
\Delta E_{br}=E(x_I)-E(x_O)-\Delta E_{fr}.
\end{equation}
For non-breaking wave groups, $\Delta E_{br}=0$, and the total energy loss therefore reduces to the background dissipation, $\Delta E_{fr}=E(x_I)-E(x_O)$.

In laboratory wave flumes, background dissipation can represent a substantial fraction of the total energy loss. For example, \citet{Banner2007} reported that approximately 20\% of the initial wave-group energy was dissipated over the propagation of non-breaking wave groups (see their figure~6). A comparable, though slightly smaller, fraction of about 15\% was observed by \citet{Cao2023} in experiments conducted in the same flume as the present study. Unless the flume is sufficiently wide \citep{Zhang2019}, accurate quantification of $\Delta E_{fr}$ is therefore essential for reliably estimating the breaking-induced energy dissipation, $\Delta E_{br}$. In the following, we revisit and assess the treatments of $\Delta E_{fr}$ and the corresponding estimates of $\Delta E_{br}$ adopted in D08, T10, and C23.

\subsection{\label{subsec: Existing methods} Existing methods used to quantify breaking wave energy dissipation in laboratory flumes}

\subsubsection{\label{subsec: Tian's method} Revisiting the T10 framework}
\begin{figure*}
    \centering
	\includegraphics[width=0.85\columnwidth]{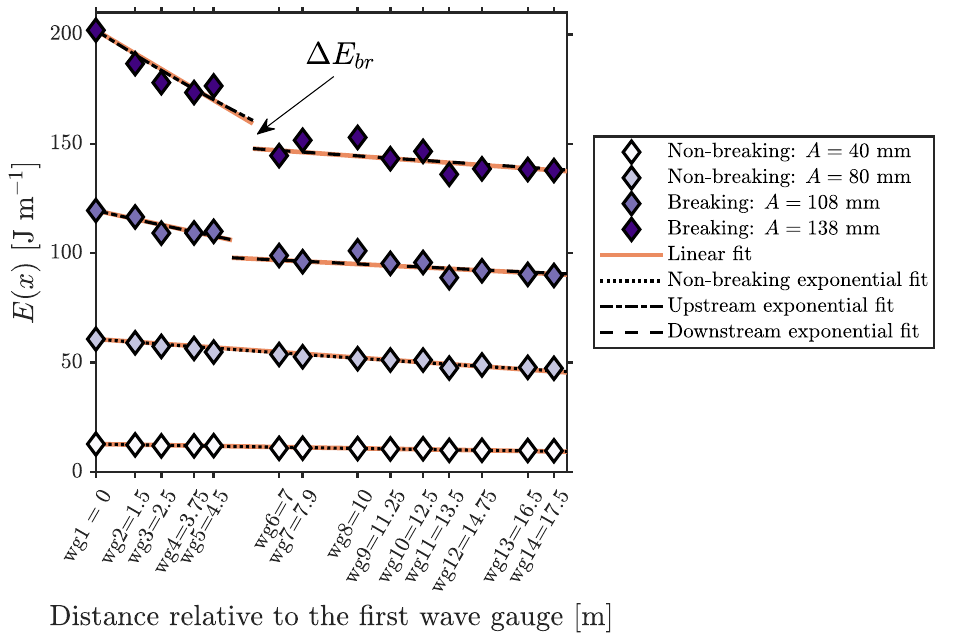}
	\caption{\label{fig: Tian method} Illustration of the T10 framework applied to two sets of \textit{SIREN} wave groups with $\gamma=2$ and $T_p=1.3$\,s, including two non-breaking and two breaking cases. The wave gauge locations are shown as relative distances from wave gauge~1. Exponential fits following equation~\eqref{eq:delta E Tian} are applied to the spatial evolution of $E(x)$. For breaking cases, separate fits are performed upstream and downstream of the identified breaking location $x_b$, and the breaking-induced energy dissipation, $\Delta E_{br}$, is evaluated from the difference between the two fits at $x_b$. For comparison, linear fits (orange lines) are also shown.}
\end{figure*}

The method proposed by \citet{Tian2010} estimates the breaking-induced energy dissipation by analysing the spatial evolution of the wave energy $E(x)$ along the flume. For non-breaking wave groups, $E(x)$ was shown to exhibit an exponential decay with distance that can be described by

\begin{equation} \label{eq:delta E Tian}
E(x) = E_0 \exp(-\sigma_T x),
\end{equation}
where $E_0\simeq E(x_I)$ and $\sigma_T$ is a spatial decay rate with units of m$^{-1}$. We illustrate the applicability of equation~\eqref{eq:delta E Tian} to non-breaking wave groups in figure~\ref{fig: Tian method}, where $E(x)$ decays in a manner that is well described by the exponential form. However, \citet{Tian2010} also noted that the exponential best fit was itself close to a straight line. For comparison, we therefore also show linear fits (orange lines) and find that, over the present measurement range, they are indeed nearly indistinguishable from the exponential fits.

When wave breaking occurs, the T10 method applies two separate exponential fits of the form given in equation~\eqref{eq:delta E Tian} to the data points upstream and downstream of an identified breaking location, $x_b$, in recognition of the different spatial decay behaviour before and after breaking. This procedure is demonstrated in figure~\ref{fig: Tian method} using two representative breaking wave groups, for which a distinct drop in $E(x)$ at $x_b$ is observed (linear fits are again shown for visual comparison and remain close to the exponential fits over the fitted ranges). The breaking-induced energy dissipation $\Delta E_{br}$ is then evaluated as the magnitude of this drop.

What becomes apparent is that only a single location $x_b$ is considered in the T10 method in inferring $\Delta E_{br}$. In practice, however, wave breaking occurs over a finite distance and the T10 method may underestimate $\Delta E_{br}$ for breaking events with a substantial streamwise extent. Also, the method requires a relatively dense spatial distribution of wave gauges in order to reliably constrain the energy decay before and after breaking; for example, the original study of \citet{Tian2010} employed approximately 30 wave gauges along the flume. Finally, measurements taken in close proximity to the breaking region are not explicitly excluded, which may introduce biases associated with strong local non-linearity that can violate the assumption of energy equipartition derived in linear theory. Because $\Delta E_{br}$ is obtained from the difference between two fitted curves at $x_b$, estimates for weakly dissipative breaking events may be sensitive to fitting uncertainty and the choice of $x_b$.

\subsubsection{\label{subsec: Drazen's method} Comparison of the methods used in D08 and C23}
\begin{figure*}
    \centering
	\includegraphics[width=0.9\columnwidth]{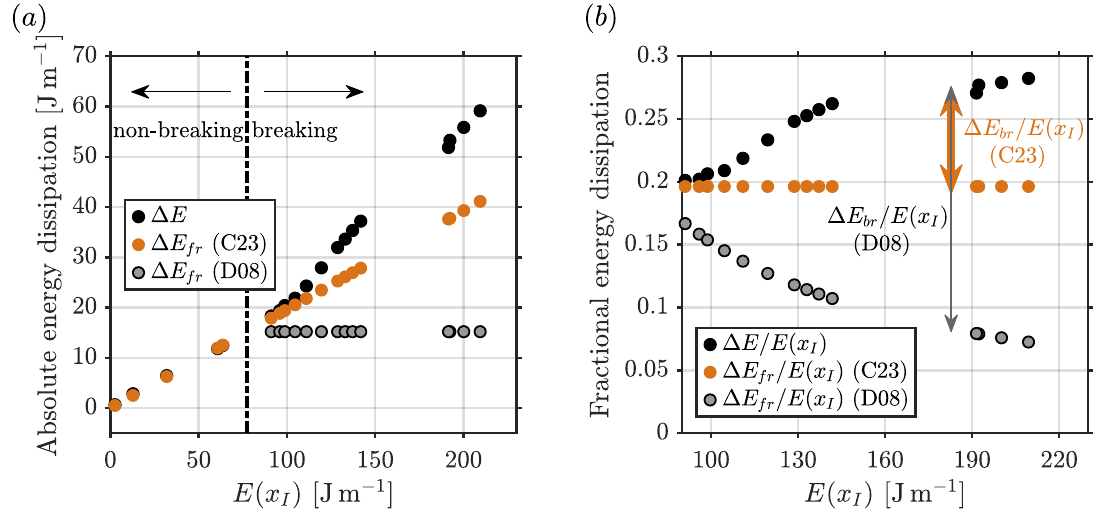}
	\caption{\label{fig: CaoDrazen} Comparison of the background energy dissipation ($\Delta E_{fr}$) and breaking-induced energy dissipation ($\Delta E_{br}$) quantified using the D08 and C23 methods for selected \textit{SIREN} wave groups with $T_p=1.3$\,s and $\gamma=2$. Panels $(a)$ and $(b)$ show the absolute and fractional energy dissipation, respectively, as functions of the upstream wave-group energy, $E(x_I)$. A range of $E(x_I)$ values between approximately 145 and 180\,J\,m$^{-1}$ corresponds to cases with multiple breaking events and is therefore absent from the analysis. In applying the D08 method, $\Delta E_{fr}$ is approximated using the value inferred from the C23 method at the average $E(x_I)$ between the most energetic non-breaking and the least energetic breaking wave groups, as indicated by the vertical dot--dashed line in panel $(a)$.}
\end{figure*}

Both D08 and C23 estimate breaking-induced energy dissipation $\Delta E_{br}$ using a control-volume-based energy balance. The key difference between the two methods lies in their respective treatments of the background energy dissipation $\Delta E_{fr}$. This difference is illustrated in figure~\ref{fig: CaoDrazen} using selected \textit{SIREN} wave groups with $T_p=1.3$\,s and $\gamma=2$.

In the D08 method, $\Delta E_{fr}$ is assumed to be a constant value and is taken as the dissipation measured for the most energetic non-breaking wave group. In the present dataset, a clearly isolated most energetic non-breaking case is not available. We therefore approximate $\Delta E_{fr}$ in D08 by taking the value inferred from the C23 method at the average initial energy between the most energetic non-breaking and the least energetic breaking wave groups, as indicated by the vertical dot--dashed line in figure~\ref{fig: CaoDrazen}$(a)$. As a result, the fractional background dissipation, $\Delta E_{fr}/E(x_I)$, given by the D08 method decreases with increasing wave group energy $E(x_I)$, as shown in figure~\ref{fig: CaoDrazen}$(b)$.

In contrast, C23 demonstrated that $\Delta E_{fr}$ scaled approximately linearly with $E(x_I)$ for their experiments, as indicated by the orange dots in figure~\ref{fig: CaoDrazen}$(a)$. This is physically consistent with increased viscous and wall-related losses for more energetic wave groups \citep{Perlin2000}. As a result, it is the fractional background dissipation, $\Delta E_{fr}/E(x_I)$, that remains approximately constant (figure~\ref{fig: CaoDrazen}$b$), and not its absolute value, $\Delta E_{fr}$. Evidence for this can also be found in the original results of \citet{Drazen2008}, whose figure~8 shows that the fractional energy loss of non-breaking wave groups remains largely unchanged with wave energy.

An underestimation of $\Delta E_{fr}$ in the D08 method consequently leads to an overestimation of the breaking-induced dissipation, $\Delta E_{br}$, as illustrated in figure~\ref{fig: CaoDrazen}$(b)$. On the other hand, C23 also noted that their formulation may slightly overestimate $\Delta E_{fr}$, and thus underestimate $\Delta E_{br}$, because $\Delta E_{fr}$ is related only to the upstream wave group energy $E(x_I)$. In practice, following breaking, a 'new' and less energetic wave group is formed downstream, which is expected to experience weaker background dissipation. This is consistent with the reduced spatial decay rates observed downstream of breaking in figure~\ref{fig: Tian method}. In the next section, we describe how this effect is accounted for and introduce the refined framework adopted in the present study.

\subsection{\label{subsec: Cao's method} A refined framework based on C23}

Notwithstanding the approximately linear relationship between the background dissipation $\Delta E_{fr}$ and the initial wave group energy reported in \citet{Cao2023} for non-breaking wave groups at a single characteristic scale ($T_p=1.2$\,s), this finding may no longer be appropriate when a broader range of underlying wave scales is considered. To verify this we quantify the total energy lost, $\Delta E$, for all non-breaking wave groups in the \textit{SIREN} dataset with $T_p$ ranging from 1.0 to 1.6\,s, and in this case $\Delta E \equiv \Delta E_{fr}$.

\begin{figure*}
    \centering
	\includegraphics[width=0.6\columnwidth]{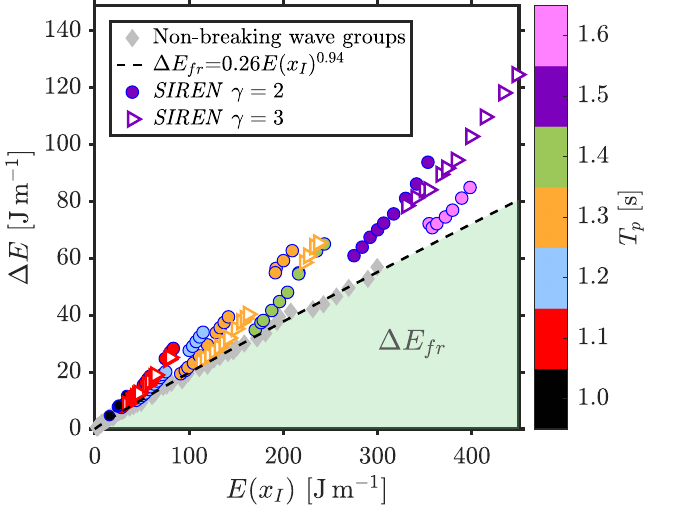}
	\caption{\label{fig: nonbreakingfit} Total energy lost over the control volume quantified for all the \textit{SIREN} wave groups plotted as a function of $E(x_I)$. Data points of different colours correspond to breaking wave groups with varying $T_p$. Non-breaking wave groups are shown as grey diamonds, while coloured symbols denote breaking cases. Fitting equation~\eqref{eq:E_{fr} Cao 1} to the non-breaking cases yields $\zeta_f = 0.26 \pm 0.02$ and $\upsilon_f = 0.94 \pm 0.015$, with a coefficient of determination $R^{2}=0.998$.}
\end{figure*}

The resulting values of $\Delta E_{fr}$ are shown in figure~\ref{fig: nonbreakingfit}, where non-breaking wave groups are indicated by grey symbols. When considered collectively across all the underlying wave scales considered here, the background dissipation is better described by a power-law relationship of the form

\begin{equation} \label{eq:E_{fr} Cao 1}
\Delta E_{fr} = \zeta_f\,E(x_I)^{\upsilon_f},
\end{equation}
where $\zeta_f$ and $\upsilon_f$ are empirical coefficients. A least-squares fit to the non-breaking wave groups shown in figure~\ref{fig: nonbreakingfit} yields $\Delta E_{fr} = 0.26(\pm0.02)\,E(x_I)^{0.94(\pm0.015)}$, where numbers in parentheses are 95\% confidence intervals on the fit coefficients. This constitutes the first modification to the C23 framework. Note that the values of $\zeta_f$ and $\upsilon_f$ are expected to depend on the specific characteristics of the wave flume (e.g. its geometry and boundary conditions) and the specific limits ($x_I$ and $x_O$) defining the control volume.

\begin{figure*}
    \centering
	\includegraphics[width=0.9\columnwidth]{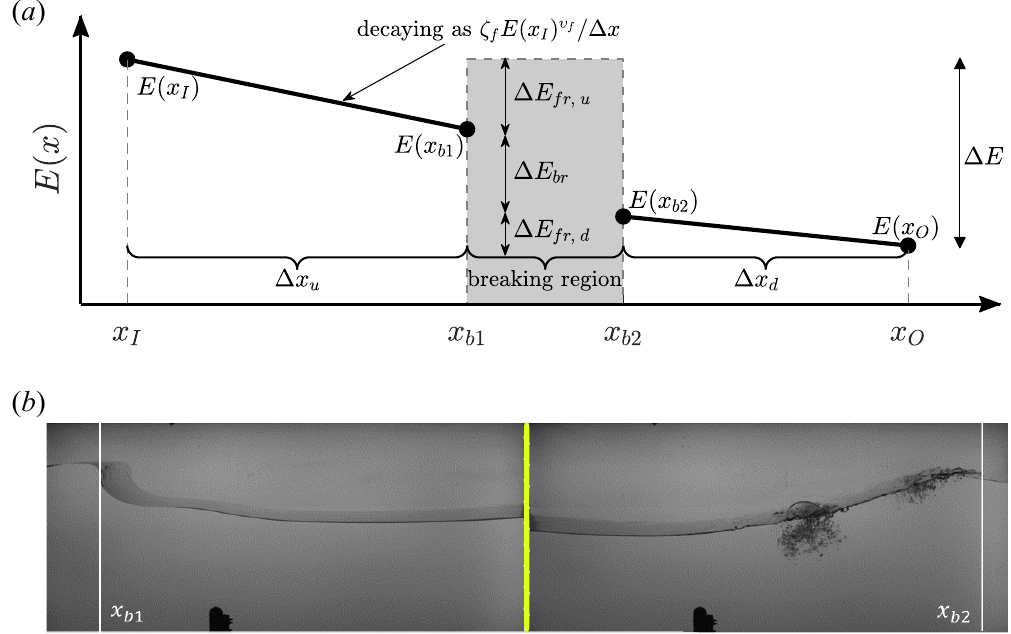}
	\caption{ \label{fig: C26 Demo} ($a$) Schematic illustration of the evolution of the wave energy $E(x)$ within the control volume in the presence of wave breaking. $E(x_{b1})$ and $E(x_{b2})$ denote the wave-group energy at the locations of incipient breaking ($x_{b1}$) and at the end of the breaking region ($x_{b2}$), respectively. The control volume is divided into three streamwise segments: an upstream region of length $\Delta x_u$, the breaking region between $x_{b1}$ and $x_{b2}$, and a downstream region of length $\Delta x_d$. The total energy loss over the control volume, $\Delta E$, is decomposed into upstream background dissipation $\Delta E_{fr,u}$, downstream background dissipation $\Delta E_{fr,d}$, and breaking-induced dissipation $\Delta E_{br}$, such that $\Delta E = \Delta E_{fr,u} + \Delta E_{fr,d} + \Delta E_{br}$. The background energy dissipation implied by equation~\eqref{eq:E_{fr} Cao 1} per unit length is indicated schematically as $\zeta_f E(x_I)^{\upsilon_f}/\Delta x$ upstream of breaking, as explained in \eqref{eq:dEdx}. ($b$) Snapshots of a single breaking event at (left) incipient breaking and (right) the end of breaking determined by hydrophone measurements. The vertical white lines indicate the locations of $x_{b1}$ and $x_{b2}$.}
\end{figure*}

Having established this improved semi-empirical method for estimating the total background energy dissipation in non-breaking wave groups, we now want to implement this to obtain a better estimate of the breaking-induced dissipation, $\Delta E_{br}$, through a segmented analysis within the defined control volume. A schematic illustration of this framework is provided in figure~\ref{fig: C26 Demo}$(a)$, and it is achieved as follows. 

We first consider a wave group entering the control volume at $x_I$ with an initial energy $E(x_I)$. As the wave group propagates through the control volume, its energy decreases due to background dissipation, as observed for non-breaking wave groups \citep{Tian2008, Tian2010} and confirmed in the present experiments (see figure~\ref{fig: Tian method} for both of our breaking and non-breaking cases). Given that equation~\eqref{eq:E_{fr} Cao 1} represents the total background energy dissipation over the full control volume, we define a constant background energy loss per unit length, which leads to the following differential form for the spatial change of the wave group energy:

\begin{equation} \label{eq:dEdx}
\frac{\mathrm{d}E}{\mathrm{d}x}
=
-\frac{\zeta_f}{\Delta x}\,E(x_I)^{\upsilon_f},
\end{equation}
where $\Delta x$ is the streamwise length of the control volume.

By doing so, the total background dissipation upstream of breaking can be quantified following \eqref{eq:dEdx} as $\Delta E_{fr,u}=\left[\zeta_f E(x_I)^{\upsilon_f}/\Delta x\right]\Delta x_u$, where $\Delta x_u=x_{b1}-x_I$ is the upstream distance over which background dissipation acts prior to breaking, such that the wave group energy at the location of incipient breaking becomes

\begin{equation} \label{eq:E_xb1_linear}
E(x_{b1}) = E(x_I)-\Delta E_{fr,u} = E(x_I) \left(1-\frac{\zeta_f}{\Delta x}E(x_I)^{\upsilon_f-1}\,\Delta x_u\right)
\end{equation}

As depicted in figure~\ref{fig: C26 Demo}$(a)$, post the breaking process, a truncated, non-breaking wave group is formed downstream of the breaking region, with an energy level $E(x_{b2})<E(x_{b1})$ at the end of breaking ($x_{b2}$). Because the location $x_{b2}$ varies between breaking events and does not coincide with a fixed position of the wave gauge, direct measurements of $E(x_{b2})$ are generally not possible and thus $E(x_{b2})$ remains unknown. This is unlike the upstream segment, where the entering energy $E(x_I)$ is explicitly measured by wave gauges and can therefore be used directly via equation~\eqref{eq:dEdx} to quantify the upstream background dissipation. An explicit estimate of $E(x_{b2})$ is therefore required in order to determine both the downstream background dissipation and the breaking-induced energy dissipation. To achieve this, and to obtain a closed expression for $E(x_{b2})$ from known quantities, we revert, for the downstream segment, to the underlying variable-dependent power law form implied by equation~\eqref{eq:dEdx}, and apply separation of variables between $x_{b2}$ and $x_O$ to infer $E(x_{b2})$ from the measured value at $x_O$. Introducing $\widetilde{E}$ as the integration variable, to distinguish it from $E(x_{b2})$ and the energy level at the outlet of the control volume $E(x_O)$, yields

\begin{equation} \label{eq:inte dEdx}
\int_{E(x_{b2})}^{E(x_O)} \frac{1}{\widetilde{E}^{\upsilon_f}}\,\mathrm{d}\widetilde{E}
=
-\frac{\zeta_f}{\Delta x}\int_{x_{b2}}^{x_O}\mathrm{d}x
=
-\frac{\zeta_f}{\Delta x}\Delta x_d,
\end{equation}
which then leads to

\begin{equation} \label{eq:E_xb2_exact}
E(x_{b2})
=
\left[
E(x_O)^{1-\upsilon_f}
+
(1-\upsilon_f)\frac{\zeta_f}{\Delta x}\,\Delta x_d
\right]^{ \frac{1}{1-\upsilon_f}},
\end{equation}
where $\Delta x_d=x_O-x_{b2}$ is the downstream distance over which background dissipation acts after breaking. The downstream background dissipation $\Delta E_{fr,d}$ is then obtained simply as the difference between the inferred energy level $E(x_{b2})$ given by \eqref{eq:E_xb2_exact} and the energy level measured at the outlet $E(x_O)$.

With the background dissipation decomposed into upstream and downstream contributions (i.e. $\Delta E_{fr}=\Delta E_{fr,u}+\Delta E_{fr,d}$; see also figure~\ref{fig: C26 Demo}$a$), the breaking-induced energy dissipation within the control volume can be calculated from the energy balance in equation~\eqref{eq:delta E br} as

\begin{subequations} \label{eq:delta E br cao}
\begin{align}
\Delta E_{br}
&= E\left( x_I\right) - E\left( x_O\right)
- \left(\Delta E_{fr,u} + \Delta E_{fr,d} \right), \label{eq:delta E br cao_a} \\
&= E(x_{b1}) - E(x_{b2}), \label{eq:delta E br cao_b} \\
&= E(x_I)\left(1-\frac{\zeta_f E(x_I)^{\upsilon_f-1}}{\Delta x}\,\Delta x_u\right)
-
\left[
E(x_O)^{1-\upsilon_f}
+
(1-\upsilon_f)\frac{\zeta_f}{\Delta x}\,\Delta x_d
\right]^{\frac{1}{1-\upsilon_f}}. \label{eq:delta E br cao_c}
\end{align}
\end{subequations}
In \eqref{eq:delta E br cao_b}, $E(x_{b1})$ and $E(x_{b2})$ are given by equations~\eqref{eq:E_xb1_linear} and \eqref{eq:E_xb2_exact}, respectively. The final expression \eqref{eq:delta E br cao_c} enables $\Delta E_{br}$ to be estimated using surface elevation measurements at the two fixed locations $x_I$ and $x_O$ defining the control volume, alongside the breaking locations $x_{b1}$ and $x_{b2}$ determined from side-view imagery.

Within the refined framework, the location $x_{b1}$ is defined as the point at which the wave crest first becomes vertical at incipient breaking, coinciding with the onset of imminent overturning \citep{Deike2015}. The downstream limit of the breaking region, $x_{b2}$, is defined as the location of the downstream end point of the breaking-induced disturbance when active breaking ceases. Both locations are identified manually from the side-view image sequences, with representative examples shown in the snapshots in figure~\ref{fig: C26 Demo}$(b)$. These locations are then converted into physical coordinates using the real-world conversion procedure described in \citet{Cao2025}. We assess the performance of the refined framework through direct comparison with the methods of D08, T10 and C23 in Appendix~\ref{app: Evaluating method}, and show that the maximum values of the breaking-induced fractional dissipation $\Delta E_{br}/E(x_I)$ obtained using the present framework can be about 10\% lower than those inferred using the D08 method.

It should be noted that in our refined framework the background dissipation occurring within the breaking region itself is neglected. This represents a limitation of the present framework, since this contribution cannot be quantified reliably within the highly turbulent two-phase flow of a breaking wave and, to our knowledge, no existing method provides a reliable means of doing so. Nevertheless, across all cases we studied, the maximum spatial extent of the breaking region was relatively short compared to the full scale of the control volume, $|x_{b1}-x_{b2}|_{\max}\sim\mathcal{O}(10\%)|x_I-x_O|$. We therefore retain this approximation without attempting to partition the energy loss within the breaking region further. An additional source of uncertainty when using equation~\eqref{eq:delta E br cao_c} arises from the identification of $x_{b1}$ and $x_{b2}$ which can be subjective. We conduct a sensitivity test which is presented in Appendix~\ref{app: model sensitivity test}, and the results demonstrate that reasonable variations in the selection of $x_{b1}$ and $x_{b2}$ do not critically affect the resulting estimates of $\Delta E_{br}$.

Finally, we want to remark that when wind is present, as in the \textit{EURUS} cases, an additional correction is applied before implementing the refined framework. This is necessary because wind can continuously transfer momentum and energy to the underlying wave field as it propagates through the control volume. The resulting wind-generated contribution to the measured wave energy is therefore quantified independently at each wave gauge location and removed before $\Delta E_{br}$ is evaluated. Details of this correction are provided in Appendix~\ref{app: wind correction}.

\section{\label{sec: Result} Results and discussion}

\subsection{\label{subsec: energy dissipation} Breaking-induced energy dissipation $\Delta E_{br}$}
\subsubsection{\label{subsubsec: steepness definitions} Definitions and implications of wave group steepness and local steepness}

As wave steepness $\mathcal{S}$ is an important variable controlling breaking-induced energy dissipation, we begin by clarifying the two types of wave steepness employed here to investigate the influence that wave scale and direct wind forcing have on the energy dissipated by wave breaking.

The first type of $\mathcal{S}$ is based on the spectral behaviour of the propagating wave group and is evaluated using surface elevation time-series measured at a fixed location upstream of breaking. As such, this wave group steepness characterises the state of the wave field prior to breaking and serves as a predictive metric, providing a means by which the threshold condition for breaking onset\footnotemark \footnotetext{In the present work, the breaking onset threshold refers to a (geometric, kinematic, or dynamic) limiting condition under which the least energetic breaking can be triggered, and is distinguished from incipient breaking which represents the instant at which the local crest is about to break.} and the associated crest steepness at incipient breaking may be inferred \citep{McAllister2024, Cao2025prep}. In addition, when considering oceanic breakers, this type of steepness can potentially be evaluated relatively readily using the spectral information available in the ocean wave models.

One wave group steepness measure of this type is $\mathcal{S}_n$, which is defined earlier in \eqref{eq: Sn}. The value of $\mathcal{S}_n$ is sensitive to local non-linear changes in the wave spectrum and therefore depends on the location at which it is evaluated relative to the breaking point \citep{Cao2025prep}. This motivates the use of an alternative wave group steepness measure defined in a spectrally energy-weighted manner \citep{Tian2010, Derakhti2016},

\begin{equation}
\mathcal{S}_s = k_s \sum_{i=1}^{N} a_i,
\qquad \text{where} \qquad
\begin{cases}
k_s \equiv k(f_s), & \text{from the linear dispersion relation}, \\[6pt]
f_s = \dfrac{\sum_{i=1}^{N} f_i \, a_i^2}{\sum_{i=1}^{N} a_i^2},
\end{cases}
\label{eq: Ss}
\end{equation}
because it was shown to better capture the breaking onset threshold and to be much less sensitive to the measurement location as the wave spectrum evolves non-linearly \citep{Cao2023, Cao2025prep}. In the present study both wave group steepness measures $\mathcal{S}_n$ and $\mathcal{S}_s$ are quantified at the inlet of the control volume ($x_I$) where we assume that wave group non-linearity remains sufficiently weak.

\begin{figure*}
    \centering
	\includegraphics[width=0.75\columnwidth]{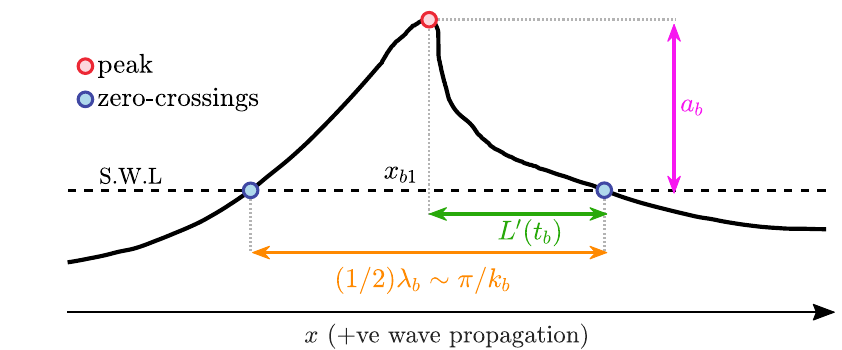}
	\caption{ \label{fig: LocalSteepness} Demonstration of the spatial parameters characterising the crest profile at incipient breaking ($t_b$). }
\end{figure*}

The second type of steepness directly quantifies the local crest geometry at incipient breaking and therefore reflects the cumulative effects of non-linear wave evolution, energy focusing and wind forcing acting up to the breaking point. This type is here referred to as local steepness. Following previous studies \citep[e.g.][]{Tian2008,Saket2017,Derakhti2018,Vita2018,Touboul2021}, the first measure of local steepness we use is defined as

\begin{equation}
\mathcal{S}_b = a_b k_b \equiv a_b \pi \left( \frac{1}{2}\lambda_b \right)^{-1},
\label{eq: Sb}
\end{equation}
where $a_b$ is the local crest amplitude at incipient breaking, $\lambda_b$ is the local wavelength estimated by doubling the zero-crossing distance (see figure~\ref{fig: LocalSteepness}) from which the local wavenumber, $k_b$, is calculated in the usual way.

Several relevant studies have suggested that the steepness of the crest front provides a more effective description of both the breaking onset and the limiting local crest form immediately prior to breaking \citep{Perlin2013, Mcallister2023, Cao2025prep}. Accordingly, we employ a second local steepness measure defined at incipient breaking as

\begin{equation}
\mathcal{S}_{\text{front}}(t_b) = \frac{a_b}{L^{\prime}(t_b)},
\label{eq: Sf}
\end{equation}
where $L^{\prime}(t_b)$ is the instantaneous horizontal length of the crest front at the time of incipient breaking $t_b$, as is illustrated in figure~\ref{fig: LocalSteepness}. It should be noted that other local measures, such as the maximum downward surface slope considered by \citet{Mcallister2023} (see their figure~7), have also been shown to provide effective descriptions of the limiting crest form, but the present study focuses primarily on assessing the ability of $\mathcal{S}_{b}$ and $\mathcal{S}_{\mathrm{front}}(t_b)$ to characterise breaking-induced energy dissipation.

\subsubsection{Breaking-induced fractional energy dissipation at different wave scales} \label{subsubsec: energy dissipation scale}

\begin{figure*}
    \centering
	\includegraphics[width=0.95\columnwidth]{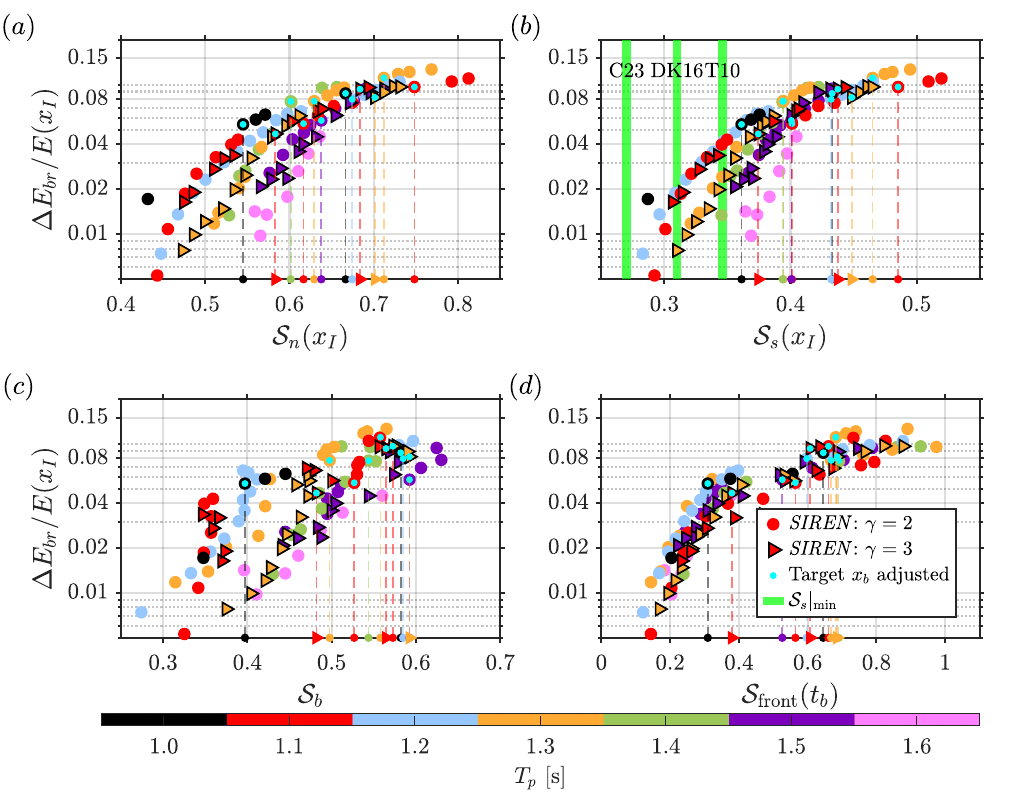}
	\caption{ \label{fig: Dissipation vs Steepness TpGamma} Breaking-induced fractional energy dissipation $\Delta E_{br}/E(x_I)$ plotted as functions of various measures of wave (group) steepness for \textit{SIREN} breaking waves: panels ($a$, $b$) use spectral measures representing wave group steepness and panels ($c$, $d$) use two locally-defined steepness measures at incipient breaking. Vertical green lines in ($b$) indicate the breaking-onset threshold steepness identified by C23 \citep{Cao2023}, T10 \citep{Tian2010}, and DK16 \citep{Derakhti2016}. The colourbar indicates different values of $T_p$. Data points marked by cyan dots correspond to cases in which the wave that actually breaks differs from the target focused wave within the group. In these cases, the target breaking location $x_b$ is adjusted to ensure that the individual breaking event occurs within the camera measurement region. The dashed lines connect these data points to their corresponding $\mathcal{S}$ values on the abscissa.}
\end{figure*}

The energy dissipation of breaking waves with varying underlying scales is examined here by plotting $\Delta E_{br}/E(x_I)$ against different measures of wave group steepness and local steepness defined in equations~\eqref{eq: Sn}, \eqref{eq: Ss}, \eqref{eq: Sb} and \eqref{eq: Sf}. The results are presented in figure~\ref{fig: Dissipation vs Steepness TpGamma} for \textit{SIREN} (unforced) breaking wave groups with varying peak periods $T_p$ (and $\gamma$), for which $\Delta E_{br}/E(x_I)$ is found to lie in the range of approximately $0.01$--$0.13$.

The dependence between $\Delta E_{br}/E(x_I)$ and wave group steepness measures, $\mathcal{S}_n(x_I)$ and $\mathcal{S}_s(x_I)$, is found to be similar (figures~\ref{fig: Dissipation vs Steepness TpGamma}$a,b$): increasing $T_p$ is associated with both a higher steepness threshold for breaking onset and a steeper increase of $\Delta E_{br}/E(x_I)$ with increasing wave group steepness. The combined effect of these two trends makes the influence of wave scale on energy dissipation difficult to characterise at any single wave group steepness value. In previous work, \citet{Cao2023} demonstrated that the influence of spectral bandwidth on energy dissipation could be largely mitigated by using the spectrally-weighted steepness $\mathcal{S}_s(x_I)$, arguing explicitly that $\mathcal{S}_s(x_I)$ unifies the breaking-onset condition across different bandwidths (see the green vertical lines in figure~\ref{fig: Dissipation vs Steepness TpGamma}$b$). This is, however, not observed here for breaking waves with different $T_p$, indicating that wave group steepness alone is insufficient to characterise breaking-induced energy dissipation across different wave scales (even when $\Delta E_{br}$ is normalised by the incoming wave group energy). This highlights the limitations in extrapolating results derived at a given scale to a larger range of scales and supports the investigation of a range of scales where practically possible.

This limitation is not removed either when using the local steepness based on zero-crossing analysis $\mathcal{S}_b$ (cf. \eqref{eq: Sb}), as shown in figure~\ref{fig: Dissipation vs Steepness TpGamma}($c$). Results corresponding to different wave scales do not collapse, and for a given value of $\mathcal{S}_b$ at individual wave scales there exists a wide range of $\Delta E_{br}/E(x_I)$, indicating the limitations of $\mathcal{S}_b$ as a measure of fractional energy dissipation in these breaking waves.

By contrast, a substantial reduction in the scatter, and hence in the apparent wave-scale dependence, is obtained when we use the crest-front steepness $\mathcal{S}_{\text{front}}(t_b)$ (figure~\ref{fig: Dissipation vs Steepness TpGamma}$d$), as defined by \eqref{eq: Sf}, to characterise the fractional energy dissipation. This can be attributed in large part to the fact that among the steepness measures examined, $\mathcal{S}_{\text{front}}(t_b)$ is more effective at capturing variations in the limiting crest geometry associated with breaking onset thresholds for (unforced) waves \citep{Cao2025prep}.

As in previous studies \citep[e.g.][]{Rapp1990,Cao2023}, we note also in our present experiments that in dispersively-focused wave groups continuing to increase the focusing amplitude $A$ may lead to breaking events with multiple breaking waves, before an isolated breaking event can be reproduced with additional increase in $A$. In the latter case, the wave that actually breaks may differ from the target focused wave within the wave group. In the present study, only isolated breaking events are retained, and cases for which a shift in the breaking wave occurs are marked in figure~\ref{fig: Dissipation vs Steepness TpGamma} using cyan dots. This shift may contribute, at least in part, to the appearance of two data clusters separated at approximately $\mathcal{S}_{\text{front}}(t_b) \approx 0.5$ in figure~\ref{fig: Dissipation vs Steepness TpGamma}($d$).

\subsubsection{Breaking-induced fractional energy dissipation at different co-flowing wind speeds} \label{subsubsec: energy dissipation wind}
\begin{figure*}
    \centering
	\includegraphics[width=0.95\columnwidth]{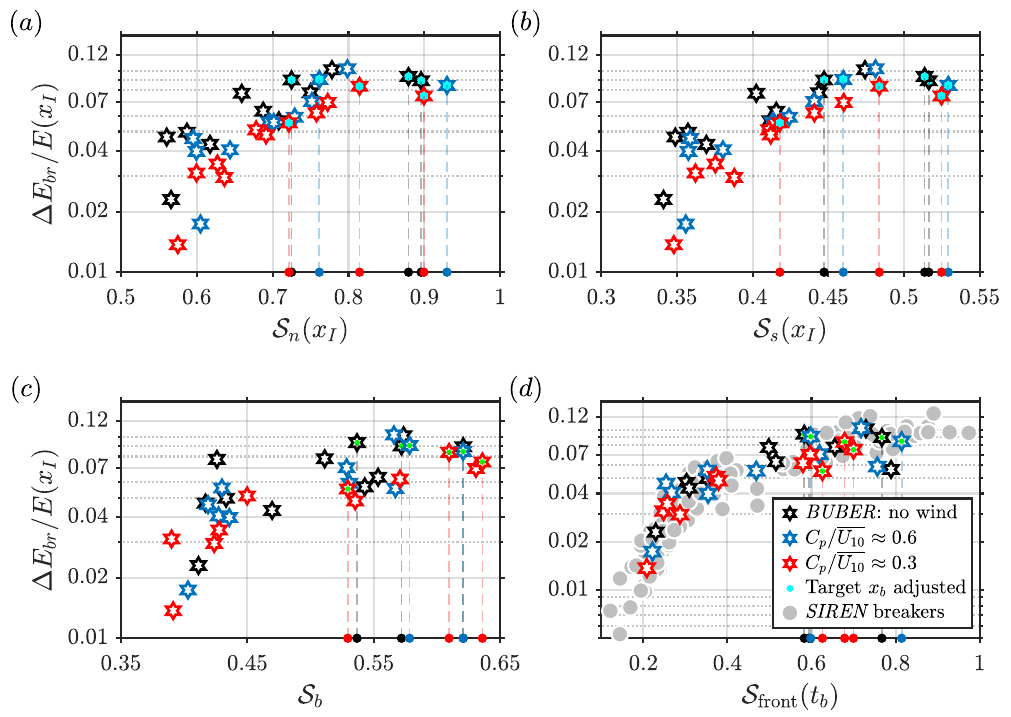}
	\caption{ \label{fig: Dissipation vs Steepness Wind} As in figure~\ref{fig: Dissipation vs Steepness TpGamma} but now for breaking waves from \textit{BUBER} (no wind) and \textit{EURUS} ($C_p/\overline{U_{10}}\approx0.6$ and 0.3) datasets. Results from \textit{SIREN} are shown as background data in ($d$) for comparison. }
\end{figure*}

We now explore the influence that co-flowing wind has on breaking-induced energy dissipation. We recall that the wave group steepness measures, $\mathcal{S}_n(x_I)$ and $\mathcal{S}_s(x_I)$, are evaluated upstream of the breaking region at small fetch and therefore remain largely independent of the applied wind forcing. As such, these steepness measures provide a reasonable reference for isolating wind effects on $\Delta E_{br}/E(x_I)$ when comparisons are made with the corresponding unforced cases.

Figure~\ref{fig: Dissipation vs Steepness Wind} presents the relationship between $\Delta E_{br}/E(x_I)$ and wave steepness under different wind conditions using data from the \textit{BUBER} (no wind) and \textit{EURUS} (characterised by wave ages $C_p/\overline{U_{10}}\approx0.6$ and $0.3$) campaigns. A key point to note first is that for a given wave group steepness, the presence of wind tends to lower the fractional energy dissipation of individual breaking waves (figures~\ref{fig: Dissipation vs Steepness Wind}$a$ and $b$). This reduction becomes more pronounced as the wind speed increases (corresponding to decreasing $C_p/\overline{U_{10}}$). This also appears to be true when the local steepness $\mathcal{S}_b$ is considered (figure~\ref{fig: Dissipation vs Steepness Wind}$c$).

There is no single mechanism through which the wind-induced reduction in $\Delta E_{br}/E(x_I)$ can be explained. Instead, the observed decrease may arise from a combined consequence of wind-induced changes in wave dispersion, high-frequency spectral energy content, and aerodynamic sheltering effects \citep{Cao2025prep}, all of which influence how waves break. From the perspective of crest-front geometry at incipient breaking, however, recent observations by \citet{Cao2025prep} indicate that wind-forced breaking tends to occur at smaller crest-front steepness values due to reduced forward leaning of the crest (n.b.\ this is consistent with the earlier breaking inception inferred by \citet{Boettger2024} based on energetic arguments). One may therefore expect part of the apparent wind-induced reduction in $\Delta E_{br}/E(x_I)$ to be compensated when the local crest-front shape is taken into account. Indeed, when we use $\mathcal{S}_{\text{front}}(t_b)$ the influence of wind forcing is largely removed (figure~\ref{fig: Dissipation vs Steepness Wind}$d$), and the wind-forced cases follow trends that are consistent with the wind-unforced data (including those from \textit{SIREN} shown in the background in figure~\ref{fig: Dissipation vs Steepness Wind}$d$ as grey circles).

\subsubsection{Subsection summary} \label{subsubsec: Summary}
Taken together, we may conclude so far from \S\ref{subsubsec: energy dissipation scale} and \S\ref{subsubsec: energy dissipation wind} that the way in which wave scale and direct wind forcing influence breaking-induced energy dissipation can be understood in terms of their respective effects on the breaking onset threshold (determined here from the limiting wave-group steepness associated with the least energetic breaking event) and on the local crest geometry at incipient breaking. Importantly, both effects are largely captured when the crest-front steepness $\mathcal{S}_{\text{front}}(t_b)$ is employed.

This finding is consistent with the working hypothesis put forward in \citet{Cao2025prep}, namely that when using wave steepness as the only controlling variable, $\mathcal{S}_{\text{front}}(t_b)$ provides the most effective description of fractional breaking-induced energy dissipation among the wave steepness measures examined. In the following sections where we examine the energy dissipation rate this finding will be used to help develop a scaling for $\Delta E_{br}/E(x_I)$ (\S\ref{subsec: dissipation scaling}). 

Regarding its application, we emphasise that $\mathcal{S}_{\mathrm{front}}(t_b)$ is purely a local diagnostic of the crest geometry at incipient breaking; although it is related to the predictive wave group steepness measures, their relationship is not described by a unique, scale-independent mapping \citep{Cao2025prep}. Therefore, the $\mathcal{S}_{\mathrm{front}}(t_b)$-based scaling that will be derived is not intended for direct implementation as a forecasting parameterisation in open-ocean spectral wave models. Instead, it is the local physical control on breaking energetics that can be evaluated from high resolution observations of individual breaking events such as stereo imaging in the open ocean. Additionally, it can be used with fully non-linear phase-resolved potential flow numerical simulations such as those carried out with a Boundary Element Method \citep[e.g.][]{Grilli1989,Grilli2001} which can capture wave evolution up to incipient breaking, but not beyond.

\subsection{\label{subsec: dissipation rate} Energy dissipation rate $\epsilon_b$}

\subsubsection{\label{subsec: tau_b} Choice of active breaking timescale and its visual interpretation}

\begin{figure*}
    \centering
	\includegraphics[width=1\columnwidth]{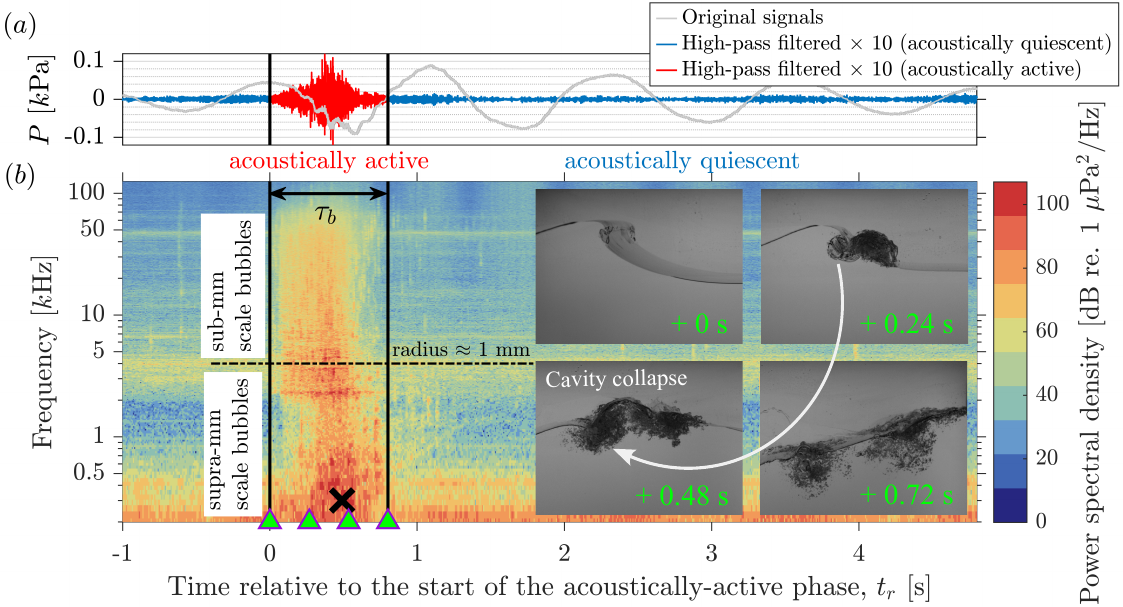}
	\caption{ \label{fig: Spectrogram} Illustration of the analysis of hydrophone acoustic outputs during a typical breaking event from the \textit{SIREN} dataset ($\gamma=3$, $T_p=1.2$~s, $A=110$~mm). ($a$) Original pressure time history and the corresponding signal after applying a 200~Hz high-pass filter (scaled by a factor of 10 for visual comparison). The region enclosed by the black lines denotes the acoustically-active phase with duration $\tau_b$. ($b$) Spectrogram calculated from the high-pass filtered signals, with colour indicating the power spectral density (PSD) in dB referenced to 1~$\mu$Pa$^{2}$\,Hz$^{-1}$. Four green triangles along the time (horizontal) axis indicate the (equally spaced) time sequence of the example images shown on the right, taken over the duration of active breaking. The black cross in the spectrogram marks the time and frequency at which the acoustic signal near 300~Hz reaches a maximum, corresponding to the breakup of the primary cavity (air pocket) entrained \citep{Deane2002, Gao2021}. The horizontal dash-dotted line denotes the characteristic frequency associated with bubbles of radius 1~mm, separating the regimes dominated by supra-mm and sub-mm bubble activities.}
\end{figure*}

As introduced in \S\ref{subsec: measurements}, we quantify the rate at which wave energy is dissipated by breaking per unit crest length as

\begin{equation}
\epsilon_b = \frac{\Delta E_{br}}{\tau_b},
\label{eq: eb}
\end{equation}
where $\tau_b$ is the duration of acoustically-active breaking inferred from hydrophone measurements. Following the procedure described in \citet{Cao2026Representative}, we process the raw hydrophone voltage signals using a 200~Hz high-pass filter to isolate the acoustic emissions associated with bubble entrainment and break-up. Figure~\ref{fig: Spectrogram}($a$) shows an example of the original and filtered pressure time histories recorded during a typical breaking event.

The associated spectrogram calculated from the high-pass filtered signal is given in figure~\ref{fig: Spectrogram}($b$), in which the duration of the acoustically-active breaking $\tau_b$ can be clearly identified together with the frequency band over which acoustic emissions occur. We may divide the latter into regimes associated with the activity of supra-mm and sub-mm bubbles, respectively \citep{Deane2002, Gao2021}, as shown by the horizontal dash-dotted line in figure~\ref{fig: Spectrogram}($b$).

To aid visual interpretation of the acoustically-active phase, we also include in figure~\ref{fig: Spectrogram}($b$) a sequence of images with the sampling times indicated by the green triangles along the time axis. At the beginning of the acoustically-active phase ($t_r=0$~s), the wave crest has already overturned and initial air entrainment is observed. This occurs later than the instant at which the crest front first becomes vertical ($t_b$), which we use to define incipient breaking, but before the jet impact. Between $t_r=0.24$~s and $t_r=0.48$~s, a pronounced peak is seen near 300~Hz in the spectrogram (marked by the black cross), corresponding to the collapse of the primary cavity where the air pocket entrained is fragmented into smaller bubbles following a cascade process \citep{Deane2002, Gao2021, Liu2024, Qi2024}. Towards the end of the acoustically-active phase ($t_r=0.72$~s) the submerged bubble plume begins to disperse, air entrainment ceases and surface disturbances diminish. 

It is worth noting that the timescale $\tau_b$ defined in the above manner is shorter than breaking durations typically inferred from side-view imagery based on visual assessment of the breaking process \citep{Tian2010, Craciunescu2020, Cao2023}, and longer than durations inferred from the growth of whitecap (foam) area in top-view imagery \citep{Deane2016, Callaghan2016} by about 15\% \citep{Cao2026Representative}. The energy dissipation rate calculated using equation~\eqref{eq: eb} should therefore be interpreted as a time average over the acoustically active phase of breaking. We adopt this definition primarily because this phase represents the stage of two-phase flow evolution during which turbulence is sufficiently intense to drive bubble fragmentation \citep{Deane2002}. Using this acoustically based definition of timescale also provides a sufficiently objective measure of the active breaking duration: identifying the end of active breaking from imagery requires a subjective assessment of when surface disturbances have subsided, particularly for highly energetic plunging events, and rising bubble plumes can produce surface agitation that further complicates this assessment.

We acknowledge that the use of $\tau_b$ may exclude the earliest part of crest overturning, between the initial fall of the crest and the onset of air entrainment, as illustrated in figure~\ref{fig: Spectrogram}($b$). However, the full overturning phase in laboratory scale breakers has a characteristic duration of only 0.1~s or so \citep{Tian2012,Cao2024}, of which the excluded interval represents only a fraction. Moreover, the energy dissipation during this early interval is expected to occur primarily through vorticity generation and associated vortical motions. Although its magnitude cannot be quantified directly using the present measurements, it is expected to be much smaller than the dissipation associated with turbulent motions and the work done against buoyancy during air entrainment after the two-phase flow has formed \citep{Derakhti2014, Callaghan2016}, which is the focus of the present study.


\subsubsection{Energy dissipation rates at varying breaking conditions}

\begin{figure*}
    \centering
	\includegraphics[width=0.95\columnwidth]{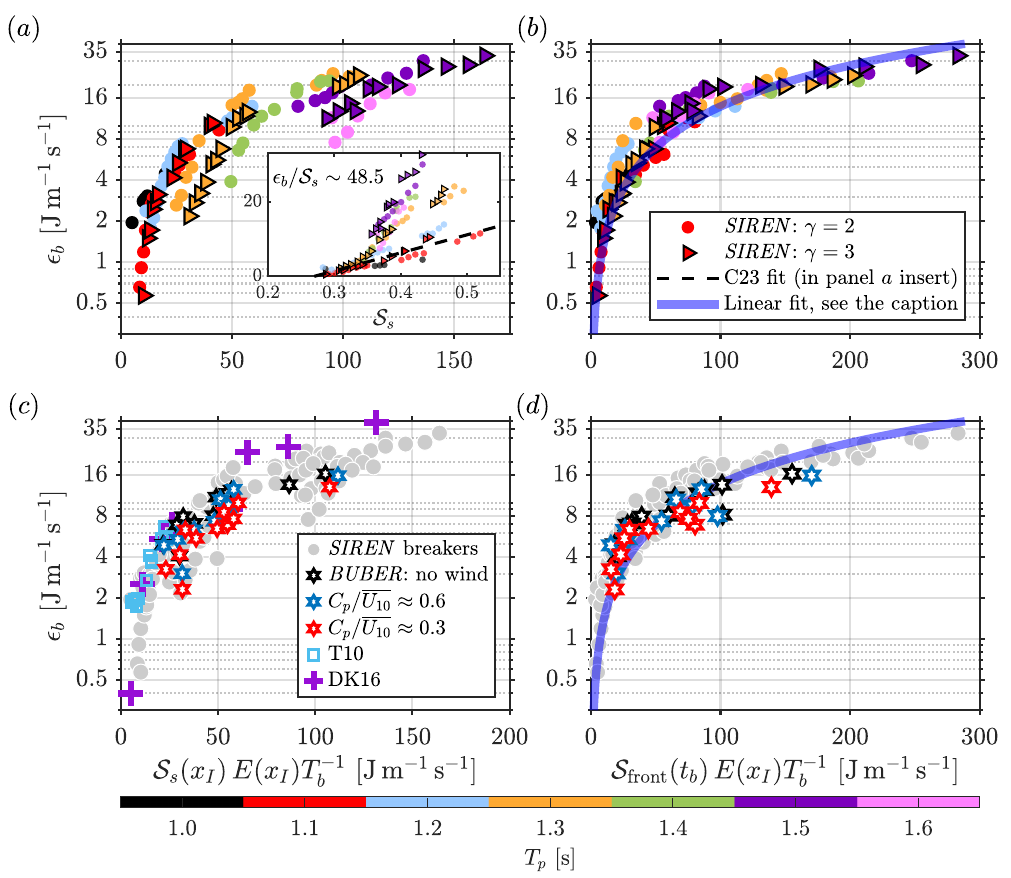}
	\caption{ \label{fig: DissipationRate vs Steepness} Energy dissipation rates of individual breaking waves $\epsilon_b$ plotted as a function of two $\mathcal{S}$ measures where ($a$, $c$) $\mathcal{S}_s(x_I)$ and ($b$, $d$) $\mathcal{S}_{\text{front}}(t_b)$. In all panels, the steepness measures are rescaled by $E(x_I)\,T_b^{-1}$. Data from \textit{SIREN} are shown in panels ($a$, $b$) and are included as background points in panels ($c$, $d$), where additional results from \textit{BUBER}, \textit{EURUS}, T10 \citep{Tian2010}, and DK16 \citep{Derakhti2016} are presented. The inset in ($a$) plots the values of $\epsilon_b$ with respect to $\mathcal{S}_{s}(x_I)$ without rescaling. Also shown is the empirical linear relationship reported by \citet{Cao2023}, who obtained a gradient of $\epsilon_b/\mathcal{S}_s \sim 48.5$ for $T_p = 1.2$~s breaking waves. This appears, however, to align more closely with the $T_p = 1.1$~s cases shown here (red dots), reflecting the slightly different methods used to estimate energy dissipation (see appendix~\ref{app: Evaluating method}). The blue lines in panels ($b$, $d$) denote linear best-fit relations given in \eqref{eq: eb fit}. The colourbar indicates different values of $T_p$. }
\end{figure*}

In figure~\ref{fig: DissipationRate vs Steepness}, values of the breaking-induced energy dissipation rate $\epsilon_b$ computed using \eqref{eq: eb} are shown. Here, we focus on the wave group steepness $\mathcal{S}_s(x_I)$ and the crest-front steepness $\mathcal{S}_{\text{front}}(t_b)$, as these two measures were shown before in \citet{Cao2023} and in \S\ref{subsubsec: energy dissipation scale}-\ref{subsubsec: energy dissipation wind} to provide a more consistent characterisation of breaking-induced energy dissipation. Since $\epsilon_b$ is a dimensional variable, an inherent dependence on the underlying wave scale can therefore be expected, as illustrated by the inset in figure~\ref{fig: DissipationRate vs Steepness}($a$). To account for this, we rescale $\epsilon_b$ by $E(x_I)^{-1}\,T_b$ but combine it with the steepness measures on the abscissa as independent variables. 

From the left column panels of figure~\ref{fig: DissipationRate vs Steepness}, $\epsilon_b$ is seen to correlate positively with $\mathcal{S}_s(x_I)\,E(x_I)\,T_b^{-1}$ across unforced laboratory conditions (\textit{SIREN}, \textit{BUBER}) as well as data we draw from literature (T10, DK16). The remaining scatter is potentially a consequence of the fact that $\mathcal{S}_s(x_I)$ does not fully capture the onset of breaking across self-similar conditions over a range of underlying wave scales (figure~\ref{fig: DissipationRate vs Steepness}$a$). In addition, consistent with the behaviour observed for fractional energy dissipation in figure~\ref{fig: Dissipation vs Steepness Wind}, imposing wind forcing is found to reduce the energy dissipation rate for similar wave group steepness (figure~\ref{fig: DissipationRate vs Steepness}$c$).

A more striking result comes from the right column panels in figure~\ref{fig: DissipationRate vs Steepness} when we employ $\mathcal{S}_{\text{front}}(t_b)$ in place of $\mathcal{S}_s(x_I)$. In this scenario, values of $\epsilon_b$ from different wind speed and wave scale conditions appear to collapse considerably onto a common trend for which we may parameterise the dependence using a linear fit, 

\begin{equation}
\epsilon_b = 0.14\,(\pm0.006)\,\mathcal{S}_{\text{front}}(t_b)\,\frac{E(x_I)}{T_b},
\label{eq: eb fit}
\end{equation}
with a coefficient of determination $r^2 = 0.87$ (blue lines in figures~\ref{fig: DissipationRate vs Steepness}$b$ and \ref{fig: DissipationRate vs Steepness}$d$). To reiterate, the use of $\mathcal{S}_{\text{front}}(t_b)$ in place of $\mathcal{S}_s(x_I)$ largely removes the underlying wave scale signal that is so clearly evident in figures~\ref{fig: DissipationRate vs Steepness}($a$) and ($c$). The above results indicate that once appropriately scaled the energy dissipation rate can be constrained effectively by the crest-front steepness, over the relatively wide range of breaking waves studied here.

\subsection{\label{subsec: dissipation scaling} Discussion on the implications of $\mathcal{S}_{\text{front}}(t_b)$ and \eqref{eq: eb fit} for constraining $\Delta E_{br}/E_0$}

As discussed in \S\ref{subsubsec: Summary}, among the various steepness measures considered, the crest-front steepness $\mathcal{S}_{\text{front}}(t_b)$ provides the most effective description of breaking-induced fractional energy dissipation. This success is achieved in part because the breaking onset threshold associated with the use of $\mathcal{S}_{\text{front}}(t_b)$ is largely scale independent, in stark contrast to the other wave steepness measures we have considered (as seen in figure~\ref{fig: Dissipation vs Steepness Wind}). 

For $\Delta E_{br}/E(x_I)$ to be parameterised by $\mathcal{S}_{\text{front}}(t_b)$, the threshold value associated with breaking onset must first be accounted for, and so the relevant independent variable is expressed as $\mathcal{S}_{\text{front}}(t_b)-\mathcal{S}_{\text{front}}(t_b)|_{\text{onset}}$. Applying this to the data in figure~\ref{fig: Dissipation vs Steepness Wind}($d$) gives $\Delta E_{br}/E(x_I) = 0.13\,(\pm0.01)\,\big[\mathcal{S}_{\text{front}}(t_b)-\mathcal{S}_{\text{front}}(t_b)|_{\text{onset}}\big]$,
with $\mathcal{S}_{\text{front}}(t_b)|_{\text{onset}} = 0.05\,(\pm0.03)$ and $r^2=0.84$. This inclusion of the onset threshold follows the same rationale as the threshold corrections applied by \citet{Romero2012} and \citet{Sinnis2021} to steepness-based predictions of the breaking strength parameter.


\begin{figure*}
    \centering
	\includegraphics[width=0.55\columnwidth]{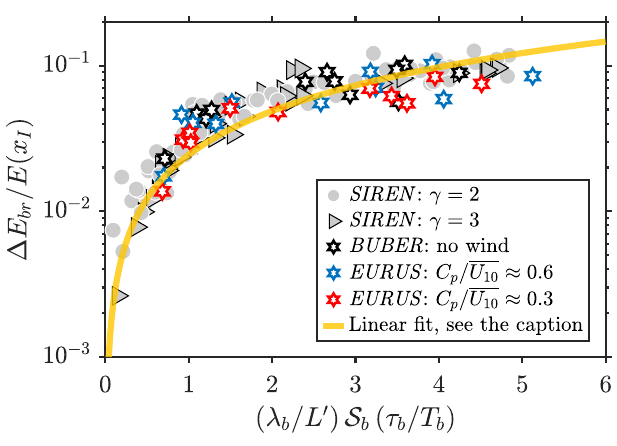}
	\caption{ \label{fig: Dissipation scaling} Comparison between fractional energy dissipation and the scaling set $(\lambda_b/L^{\prime}) \, \mathcal{S}_b \, (\tau_b/T_b)$ for all breaking waves from the \textit{SIREN}, \textit{BUBER} and \textit{EURUS} datasets. The yellow line shows the linear best fit forced through the origin, with a gradient of $0.024 (\pm0.001)$ and $r^2=0.8$. The fitted gradient is consistent with the value of $0.14/(2\pi)$ given by equation~\eqref{eq: eb fit}.}
\end{figure*}

Alternatively, further insight into the controls on fractional energy dissipation can be obtained from the collapse of the dissipation rate measurements shown in figure~\ref{fig: DissipationRate vs Steepness}($d$). The definition of $\mathcal{S}_{\text{front}}(t_b)$ in \eqref{eq: Sf} can be recast in terms of the local steepness $\mathcal{S}_b$ ($=a_b k_b$) and a fore--aft asymmetry of the crest ($(k_b L^{\prime})^{-1}$). If this is combined with the definition of the energy dissipation rate in \eqref{eq: eb} and its parametrised form in \eqref{eq: eb fit}, we arrive at a more physically motivated scaling for fractional energy dissipation:

\begin{equation}
\frac{\Delta E_{br}}{E_0} \sim \frac{\Delta E_{br}}{E(x_I)} \propto \frac{2\beta^{*}}{\pi}\, \mathcal{S}_b\, \frac{\tau_b}{T_b},
\label{eq: fractional fit}
\end{equation}
where $\beta^{*}=\lambda_b/(4L^{\prime}$) is a crest-leaning parameter we define that equals unity for symmetric crests and increases as the crest leans forward in the event of breaking (i.e. $L^{\prime}$ decreases in relation to the full wavelength $\lambda_b$). To show the proportional relationship implied by equation~\eqref{eq: fractional fit} we plot $\Delta E_{br}/E(x_I)$ against $(\lambda_b/L^{\prime})\,\mathcal{S}_b\,(\tau_b/T_b)$ in figure~\ref{fig: Dissipation scaling} using all the breaking waves from the \textit{SIREN}, \textit{BUBER}, and \textit{EURUS} experiments. As expected from the above derivation, the measurements follow an approximately linear relationship across the range of breaking conditions examined. The best fit constrained through the origin (the yellow line) has a gradient of 0.024, sufficiently close to $0.14/(2\pi)$ predicted from equation~\eqref{eq: eb fit}.

Most importantly, equation~\eqref{eq: fractional fit} has broader implications for what we presently understand about how $\Delta E_{br}/E_0$ is characterised by $\mathcal{S}$-based measures. In addition to local non-linearity and crest asymmetry (which are naturally incorporated in the definition of $\mathcal{S}_{\text{front}}(t_b)$), it identifies the non-dimensional breaking duration $\tau_b/T_b$ as an additional controlling variable for fractional energy dissipation. Indeed, although assumptions have been taken to tie the (absolute) breaking duration to a characteristic wave period \citep[e.g.][]{Deike2016, Derakhti2018}, there is evidence that $\tau_b/T_b$ is not fixed but instead varies with how dissipative an individual breaking event is \citep{Perlin2013, Cao2026Representative}. On this basis, our explicit inclusion of $\tau_b/T_b$ in \eqref{eq: fractional fit} echoes these observations and enables variations in breaking duration to be incorporated into a $\mathcal{S}$-based description of $\Delta E_{br}/E_0$.

\subsection{\label{subsec: bb} The breaking strength parameter $b$}
In this final results section, we examine how the breaking strength parameter varies across different underlying wave scales and co-flowing wind conditions. Following \citet{Tian2010, Allis2013, Derakhti2016, Derakhti2018} and \citet{Craciunescu2020}, we evaluate \eqref{eq: b} by taking $c \sim c_b$, where $c_b = 2\pi/(T_b k_b)$. In other words, $c_b$ is the local phase speed determined from the zero-crossing wavelength $\lambda_b$, different from the instantaneous crest propagation speed that can experience a transient slowdown as an unsteady crest approaches its maximum steepness before breaking \citep{Banner2014crestspeed, Fedele2020}. To distinguish this particular estimate from the more general definition of $b$ in \eqref{eq: b}, we denote it hereafter as $b_b$. We then consider simple parametric forms of the type $b_b \propto \mathcal{S}^{\Xi}$ (see \eqref{eq: power-law}) and compare the resulting fits across datasets.

\begin{figure*}
	\includegraphics[width=1.0\columnwidth]{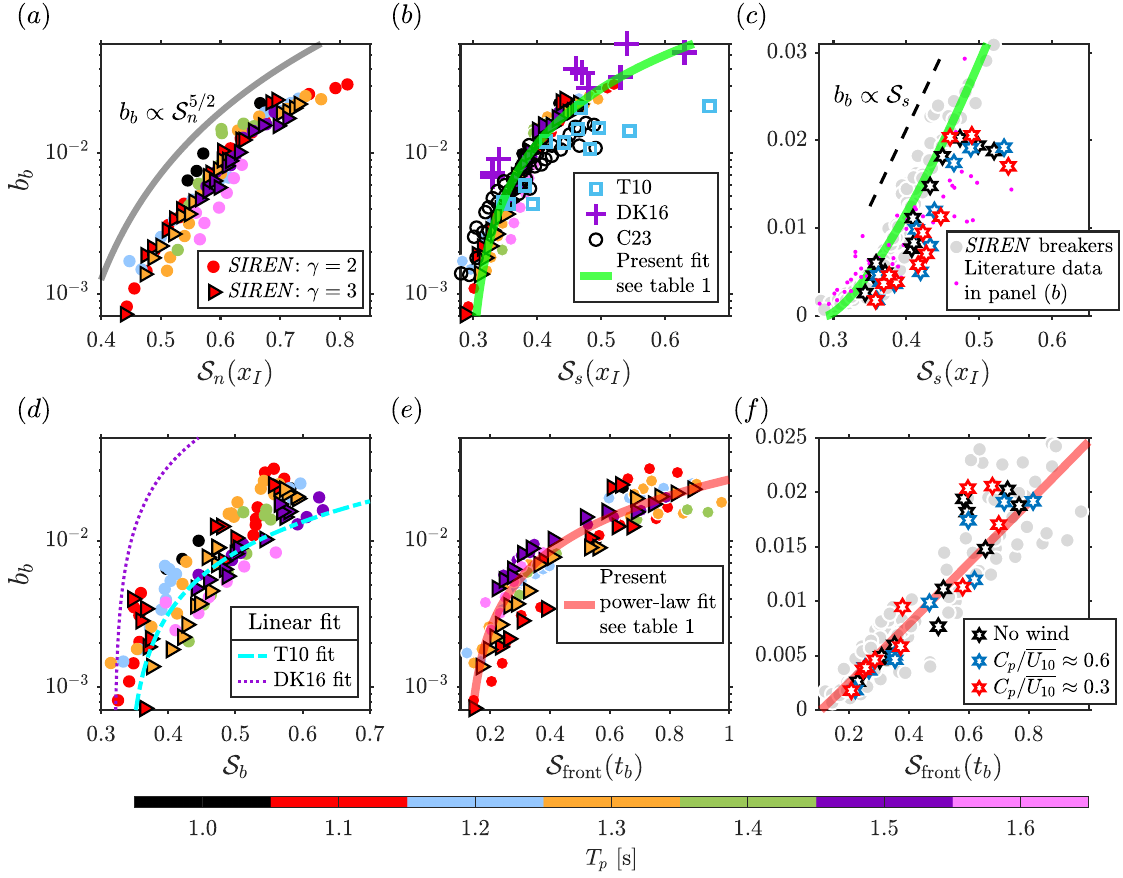}
	\caption{ \label{fig: bb vs Steepness} Relationships between the breaking strength parameter $b_b$ calculated using equation~\eqref{eq: b}, and different measures of ($a$--$c$) wave group steepness and ($d$--$f$) local steepness. Again, different values of $T_p$ are given in the colourbar. The grey line in ($a$) shows the $\mathcal{S}_{n}^{5/2}$ scaling. Results derived from T10, DK16 and C23 are shown as different markers in ($b$), together with the linear best fit applied to the present dataset (green line). In ($d$) the purple dotted and cyan solid lines show the linear $b_b$--$\mathcal{S}_b$ relationships reported respectively in DK16 and T10. The red curve in ($e$) shows the power-law based parametric fit to the breaking waves in the present study. Panels ($c$) and ($f$) replot the relationships in ($b$) and ($e$), respectively, on linear axes and add the \textit{BUBER} and \textit{EURUS} data while retaining the original data in the background for comparison; separate panels are used to avoid overcrowding. Fit coefficients for the relationships shown above are summarised in table~\ref{tab:fit_coeffs_source}.
    }
\end{figure*}

Figure~\ref{fig: bb vs Steepness} presents the behaviour of $b_b$ in relation to different measures of wave steepness. The results in figures~\ref{fig: bb vs Steepness}($a$) and ($b$) show that, at a broad level, $b_b$ for unforced (\textit{SIREN}) breaking waves can be described by previously proposed scaling relationships. These include the $\Xi=5/2$ power law $b_b \propto \mathcal{S}_n^{5/2}$ based on the inertial scaling of turbulent dissipation \citep{Drazen2008, Romero2012} (figure~\ref{fig: bb vs Steepness}$a$), and the approximately linear dependence on spectrally weighted steepness, $b_b \sim \mathcal{S}_s$, reported in earlier studies \citep{Tian2010, Derakhti2016, Cao2023} (figure~\ref{fig: bb vs Steepness}$b$). 

Note that in the latter case we relax the strict linear constraint and instead fit the present dataset using the more general threshold-type power-law form adopted in the original \cite{Cao2023} study (green line in figure~\ref{fig: bb vs Steepness}$b$, coefficients listed in table~\ref{tab:fit_coeffs_source}). Although this fit is not strictly linear, it shows reasonable agreement with the literature results of T10, DK16 and C23 (figure~\ref{fig: bb vs Steepness}$b$), and remains nearly linear over most of the observed range, particularly for $\mathcal{S}_s(x_I) \gtrsim 0.35$, as illustrated by the dashed linear reference line in figure~\ref{fig: bb vs Steepness}($c$). However, this parameterisation does not adequately reconcile the influence of wind forcing, as evidenced by the systematically lower values of $b_b$ observed under stronger wind conditions (figure~\ref{fig: bb vs Steepness}$c$). Such dependence is consistent with the results discussed earlier, where wind forcing was shown to promote earlier breaking and to limit both the energy dissipation rate and, consequently, $b_b$.

\begin{table}
\centering
\begin{tabular}{p{5.0cm}ccc c}
\toprule
Source & \multicolumn{2}{c}{Fit coefficients} & & Goodness of fit \\
\cmidrule(lr){2-4} \cmidrule(lr){5-5}

 &  & &  & \\

\multicolumn{5}{c}{\textbf{Linear fit:} $b_b=\alpha_1\,(\mathcal{S}-\alpha_2)$} \\
\midrule
 & $\alpha_1$ & $\alpha_2$ &  & $r^2$ \\
\midrule

Figure~\ref{fig: bb vs Steepness}($d$): T10 \citep{Tian2010} 
& 0.051 & 0.017 &  & --- \\

Figure~\ref{fig: bb vs Steepness}($d$): DK16 \citep{Derakhti2016} 
& 0.40 & 0.32 &  & --- \\

 &  & &  & \\

\midrule
\multicolumn{5}{c}{\textbf{Power-law fit:} $b_b=\alpha_3\,(\mathcal{S}-\alpha_4)^{\Xi}$} \\
\midrule
 & $\alpha_3$ & $\alpha_4$ & $\Xi$ & $r^2$ \\
\midrule

Figures~\ref{fig: bb vs Steepness}($b,c$): green line
& 0.260 ($\pm$ 0.094) & 0.289 ($\pm$ 0.024) & 1.41 ($\pm$ 0.30) & 0.93 \\

Figures~\ref{fig: bb vs Steepness}($e,f$): red line \eqref{eq: bb_Sfront_fit}
& 0.027 ($\pm$ 0.002) & 0.106 ($\pm$ 0.088) & 1.02 ($\pm$ 0.29) & 0.90 \\
\bottomrule
\end{tabular}
\caption{Coefficients of the different parametric fits shown in figure~\ref{fig: bb vs Steepness}. Values in parentheses denote $95\%$ confidence intervals, and the corresponding goodness-of-fit measure is $r^2$.}
\label{tab:fit_coeffs_source}
\end{table}

Whilst the spectrally-based steepness scalings discussed above are useful for characterising the overall trends and magnitude of $b_b$, implementing them for field observations of individual breaking waves or whitecaps is not straightforward \citep{Vita2018, Cao2025prep}. We also acknowledge recent promising attempts to link $b_b$ to dynamic (energetic) properties of breaking waves \citep{Derakhti2018, Boettger2024}, although these quantities are likewise non-trivial to constrain for individual events in the field. These constraints motivate us to examine local, geometry-based steepness measures that can be determined more directly from optical remote-sensing observations \citep{Callaghan2024a, Peach2025}.

In this light, we examine $b_b$ in relation to local steepness measures in figures~\ref{fig: bb vs Steepness}($d$)--($f$). Although both \citet{Tian2010} and \citet{Derakhti2016} reported linear relationships between $b_b$ and $\mathcal{S}_b$, their respective trends differ substantially and do not align well with the trend observed in the present dataset, which itself exhibits considerable scatter (figure~\ref{fig: bb vs Steepness}$d$).

When $\mathcal{S}_{\text{front}}(t_b)$ is used, however, as shown in figures~\ref{fig: bb vs Steepness}($e$) and ($f$), we obtain much improved data collapse. This is observed both across unforced cases at different wave scales (relative to figure~\ref{fig: bb vs Steepness}$d$) and under wind forced conditions (relative to figure~\ref{fig: bb vs Steepness}$c$). In the absence of an a priori constraint on the functional dependence we fit our data empirically using again a power-law form with a freely-tuned $\Xi$, yielding

\begin{equation}
b_b = 0.027 \big(\mathcal{S}_{\text{front}}(t_b)-0.106\big)^{1.02},
\label{eq: bb_Sfront_fit}
\end{equation}
where 0.106 is the fitted breaking onset threshold $\mathcal{S}_{\text{front}}(t_b)|_{\text{onset}}$ (see also table~\ref{tab:fit_coeffs_source} for further details of \eqref{eq: bb_Sfront_fit}, including the 95\% confidence intervals for the coefficients). We note that this threshold is larger than the estimate of 
$0.05\pm0.03$ obtained in \S\ref{subsec: dissipation scaling} by linearly extrapolating $\Delta E_{br}/E(x_I)$ to zero. This difference arises because the two estimates use different dependent variables and regression forms. Nonetheless, the two estimates are not statistically significantly different at the 95\% confidence level, since their respective 95\% confidence intervals, $[0.018,0.194]$ and $[0.02,0.08]$, overlap. 

The key feature of equation~\eqref{eq: bb_Sfront_fit} is its near-unity exponent, which implies an approximately linear dependence between $b_b$ and the threshold-corrected crest-front steepness, although the broader generality of this empirical relation remains to be tested.


\section{\label{sec: conclusion} Conclusions}
In this study, we have used laboratory measurements of unsteady, unidirectional surface wave groups to investigate breaking-induced energy dissipation and dissipation rates in individual breaking events, across a range of underlying wave scales and under both wind-unforced and co-flowing wind conditions.

Our first contribution is the development of a refined framework for quantifying the breaking-induced energy loss $\Delta E_{br}$ in physical wave flumes. Compared with the approaches commonly employed in the literature, our framework more reliably accounts for background dissipation arising from wave group propagation, without introducing substantial additional measurement complexity. We learn that treatments commonly used in the past where background dissipation is not explicitly separated from $\Delta E_{br}$ can lead to a systematic overestimation of $\Delta E_{br}$, by up to $\mathcal{O}(0.1)$ in terms of $\Delta E_{br}/E_0$ for the conditions considered here.

Using this framework, and by examining a range of steepness $\mathcal{S}$ measures, we find that the primary influence of wave scale on both fractional energy dissipation $\Delta E_{br}/E_0$ and dissipation rate $\epsilon_b$ is through its role in setting the breaking onset. On the other hand, both $\Delta E_{br}/E(x_I)$ and $\epsilon_b$ are found to decrease systematically in the presence of co-flowing wind. This reduction is unlikely to arise from a single mechanism, but rather from a combination of wind-related effects (e.g. modifications to wave dispersion, changes in the high-frequency spectral energy content, and alterations to particle kinematics within the crest region). It is nonetheless consistent with existing evidence that wind promotes earlier breaking and reduces the forward leaning of the crest at incipient breaking \citep{Boettger2024, Cao2025prep}.

Given its demonstrated capability among the steepness measures considered to characterise both the breaking onset and the local crest geometry at incipient breaking, the crest-front steepness $\mathcal{S}_{\text{front}}(t_b)$ provides a natural basis for examining breaking energetics. Building on this, we find that, once appropriately scaled, the energy dissipation rate $\epsilon_b$ can be effectively constrained by $\mathcal{S}_{\text{front}}(t_b)$ across all breaking conditions studied here. This then leads to a scaling for fractional energy dissipation of the form $\Delta E_{br}/E(x_I) \propto \beta^{*}\,\mathcal{S}_b\,(\tau_b/T_b)$, where $\beta^{*}$ is a crest-leaning parameter, $\mathcal{S}_b$ is the local zero-crossing steepness at incipient breaking, and $\tau_b/T_b$ is the non-dimensional breaking duration. This scaling highlights that $\Delta E_{br}/E(x_I)$ depends not only on local non-linearity, but also on crest asymmetry and the breaking duration relative to the wave period. Whilst in many studies $\tau_b/T_b$ has been implicitly assumed to be fixed, our results reinforce that it plays an active role in controlling dissipation in individual breaking waves.

Finally, we revisited the behaviour of the breaking strength parameter across a wide range of wave steepness definitions. We first confirm that the scaling laws proposed previously, including $b_b \propto \mathcal{S}_n^{5/2}$ \citep{Drazen2008, Romero2012} and $b_b \propto \mathcal{S}_s$ \citep{Tian2010, Derakhti2016, Cao2023}, capture the overall trends across much of the dataset. Motivated by practical limitations associated with applying such measures to individual breaking events in the field, we further relate $b_b$ to locally-measured $\mathcal{S}_{\text{front}}(t_b)$. We find an approximately linear relationship between $b_b$ and $\mathcal{S}_{\text{front}}(t_b)$ above the breaking onset threshold. Looking ahead, a natural next step involves examining how the relationships identified in this study extend to oceanic breaking waves where additional physical processes, such as the inherently three-dimensional effects, can come to play a part.

%

\section*{Funding}
The present work is supported by the Postdoctoral Innovation Program of Shandong Province (grant number: SDCX-ZG-202601014), the Qingdao Postdoctoral Science Foundation (grant number: QDBSH20250202010) and the China Postdoctoral Science Foundation (grant number: 2025M770858) awarded to R.C. The experimental data used here were collected during R.C.'s PhD at Imperial College London during which he was supported by the Skempton PhD Scholarship, and the experimental campaigns were funded by a NERC Grant (grant number: \href{https://gtr.ukri.org/projects?ref=NE\%2FT000309\%2F1}{NE/T000309/1}) awarded to A.H.C.

\section*{Author contributions}
R.C.: Project design, Conceptualisation, Data collection, Methodology, Data analysis, Writing - Original draft, Writing - Review \& Editing, Funding Acquisition; 
E.M.P.: Project design, Data collection, Methodology, Writing - Review \& Editing; 
X.C.: Writing - Review \& Editing, Supervision; 
A.H.C.: Project design, Conceptualisation, Writing - Review \& Editing, Supervision, Funding Acquisition, Final approval.

\section*{Declaration of interests}
The authors report no conflict of interest.

\section*{Author ORCID}
Rui Cao, https://orcid.org/0000-0002-2360-9412;  \\
Enrique M. Padilla, https://orcid.org/0000-0003-0532-8809; \\
Xu Chen, https://orcid.org/0000-0001-8526-9612;\\
Adrian H. Callaghan, https://orcid.org/0000-0001-7531-5072.

\appendix

\section{\label{app: Evaluating method} Method comparisons}
\begin{figure*}
	\includegraphics[width=1\columnwidth]{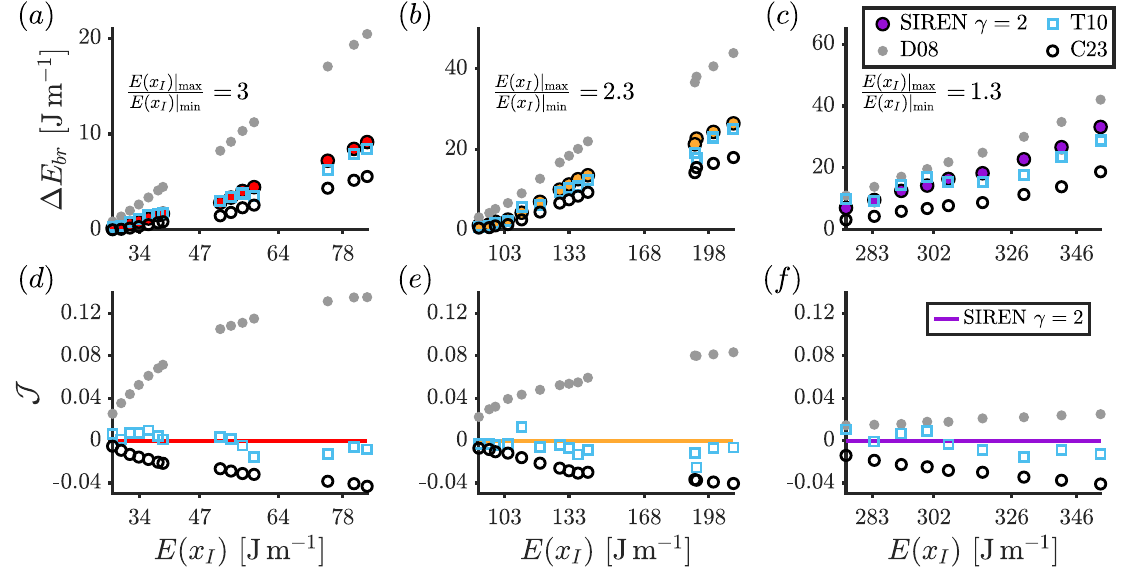}
	\caption{\label{fig: MethodComparisons_all} Comparison of calculated values of ($a$)-($c$) breaking-induced energy dissipation $\Delta E_{br}$ and ($d$)-($f$) $\mathcal{J}$ based upon different methods outlined in \S\ref{sec: method} for a selection of \textit{SIREN} breaking waves with $\gamma=2$ and $T_p=1.1$\,s (left panels), 1.3\,s (middle panels) and 1.5\,s (right panels). The values of $E(x_I)|_{\text{max}}/E(x_I)|_{\text{min}}$ displayed in ($a$)-($c$) indicate the ratio of the maximum to the minimum $E(x_I)$ for a given ($T_p$, $\gamma$) combination, highlighting how much the wave group energy has increased.}
\end{figure*}

In this appendix we compare estimates of breaking-induced energy dissipation, $\Delta E_{br}$, obtained using the methods of D08, T10, and C23, with those produced by the refined framework we propose in \S\ref{subsec: Cao's method}. For demonstration purposes results are presented for wave groups from the \textit{SIREN} dataset with $\gamma=2$ and $T_p = [1.1,\,1.3,\,1.5]$\,s, with $\Delta E_{br}$ plotted as a function of $E(x_I)$ in figures~\ref{fig: MethodComparisons_all}($a$)--($c$).

We see that across all breaking conditions examined, the D08 method tends to systematically overestimate $\Delta E_{br}$ relative to the present refined framework, whilst the C23 method underestimates it. The magnitude of this overestimation (underestimation) generally increases with increasing $E(x_I)$. Estimates obtained using the T10 method and the refined framework are, by contrast, similar and typically fall between the values given by D08 and C23. We also see a reduced difference in $\Delta E_{br}$ among the different methods at larger $T_p$ (e.g.\ figure~\ref{fig: MethodComparisons_all}$c$), and this may be associated with the more limited upper bound of $E(x_I)$ that can be achieved by larger-scale wave groups (see the decreasing ratio $E(x_I)|_{\max}/E(x_I)|_{\min}$ texted from figure~\ref{fig: MethodComparisons_all}$a$ to figure~\ref{fig: MethodComparisons_all}$c$).

In more quantitatively comparing different methods we define the dimensionless metric:

\begin{equation} \label{eq: Gamma}
\mathcal{J} = \frac{\Delta E_{br} - \Delta E_{br, \text{ calculated using the refined framework}}}{E(x_I)},
\end{equation}
which measures the differences in breaking-induced fractional energy dissipation from the value obtained using the refined framework proposed in the present study. 

The resulting values of $\mathcal{J}$ for different breaking conditions are presented in figures~\ref{fig: MethodComparisons_all}($d$)--($f$). Using this metric we find that the maximum overestimation (underestimation) associated with the D08 (C23) method at smaller $T_p$ can reach $\sim0.12$ ($\sim0.04$), which reduces significantly when either $E(x_I)$ decreases or $T_p$ increases. Consistent with the trends observed in figures~\ref{fig: MethodComparisons_all}($a$)--($c$), using the T10 method yields values of $\Delta E_{br}$ that are in close agreement with those obtained using the refined framework, with deviations typically within approximately 0.02 of the refined estimates.

\section{Sensitivity test on the refined framework} \label{app: model sensitivity test}

Within the application scope of our refined framework for estimating breaking-induced energy dissipation the manual selection of the two horizontal locations $x_{b1}$ and $x_{b2}$ (see figure~\ref{fig: C26 Demo}$b$) inevitably involves a degree of subjectivity. It is therefore important to assess how sensitive the estimated $\Delta E_{br}$ is to this choice. In this appendix, we examine this sensitivity by recalculating $\Delta E_{br}$ under deliberately imposed, large variations in the horizontal breaking distance $|x_{b1}-x_{b2}|$.

\begin{figure*}
\centering
	\includegraphics[width=0.9\columnwidth]{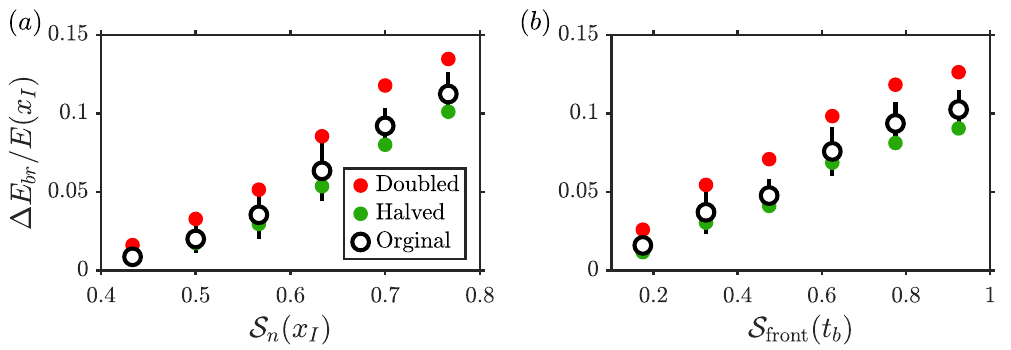}
	\caption{ \label{fig: DissipationPara} Bin-averaged values of the breaking-induced fractional energy dissipation plotted against ($a$) wave group steepness $\mathcal{S}_n(x_I)$ and ($b$) crest-front steepness at incipient breaking $\mathcal{S}_{\text{front}}(t_b)$ for bulk \textit{SIREN} breaking wave groups, calculated using the original, halved, and doubled breaking distances $|x_{b1}-x_{b2}|$. Vertical lines indicate $\pm1$ standard deviation intervals of the bin averages computed using the original $|x_{b1}-x_{b2}|$. }
\end{figure*}

Figure~\ref{fig: DissipationPara} compares the resulting fractional energy dissipation computed using halved, original, and doubled values of $|x_{b1}-x_{b2}|$, plotted against two representative measures of wave (group) steepness. The data are presented in bin-averaged form to more clearly reveal the underlying trends. As energy dissipation increases with steepness, the reduction of $|x_{b1}-x_{b2}|$ by half (green dots) leads to a slight underestimation of $\Delta E_{br}/E(x_I)$, although the resulting values remain well within the standard deviation bounds associated with the original breaking distance. Conversely, even when $|x_{b1}-x_{b2}|$ is doubled (red dots), the resulting increase in $\Delta E_{br}/E(x_I)$ reaches at most $\sim0.03$ at the highest steepness values considered.

Such large deviations in the effective breaking distance are, however, unlikely to arise in practice. The above results therefore indicate that the refined framework retains a high degree of robustness, with moderate subjectivity in the selection of $x_{b1}$ and $x_{b2}$ exerting only a limited influence on the estimated breaking-induced energy dissipation.

\section{Treatment of wind-related energy contributions in the refined framework} \label{app: wind correction}
For the wind-forced \textit{EURUS} experiments, the measured surface elevation records contain contributions from both the prescribed focused wave group and the background wind-generated wave field. If the latter were retained in the calculation of $E(x)$ via equation~\eqref{eq:eta energy}, its fetch-dependent evolution (growth) would contribute to the measured energy difference between the inlet and outlet of the control volume and could therefore lead to an underestimation of $\Delta E_{br}$. We account for this contribution by independently quantifying the wind-generated wave energy at each wave gauge location and subtracting it from the energy obtained from the mixed field comprising the wind-generated waves and the paddle-generated focused wave group.

The feasibility of this approach is supported by the measurements of wave spectra reported in \citet{Cao2025prep}. As shown in figure~17 of their Appendix~A.2, for non-linear, finite amplitude focused wave groups, the high-frequency tail of the spectrum measured under the same wind forcing considered here closely follows that of the corresponding purely wind-generated wave field, at varying fetch locations. This implies that the wind-generated short wave components in the present experiments are superimposed approximately linearly onto the paddle-generated focused wave group. We therefore use the first order spectral decomposition

\begin{equation}
S_{\eta\eta}^{m}(f,x) \simeq S_{\eta\eta}^{p}(f,x) + S_{\eta\eta}^{w}(f,x),
\label{eq:wind_spectral_decomposition}
\end{equation}
where the superscripts $m$, $p$ and $w$ represent the measured mixed field, the prescribed focused wave group and the purely wind-generated wave field, respectively. Note that equation~\eqref{eq:wind_spectral_decomposition} is used here only to separate the independently measurable background wind-wave contribution when evaluating $\Delta E_{br}$; it does not correct for any wind-induced changes in the propagation, spectral evolution or local geometry of the prescribed wave group, for which their combined influence on breaking-induced energy dissipation is the wind effect examined here.

In each \textit{EURUS} run, the wind was applied continuously before the focused wave group was generated. This gave a 256-s surface elevation record of the purely wind-generated wave field, obtained after it had reached a statistically stationary state but before the wave paddle was activated. At each wave gauge location, the corresponding variance-density spectrum, $S_{\eta\eta}^{w}(f,x)$, was estimated using 64-s Hamming windows with 75\% overlap. Its zeroth spectral moment is

\begin{equation}
m_0^{w}(x) = \int_{f_l}^{f_u} S_{\eta\eta}^{w}(f,x)\,\mathrm{d}f,
\label{eq:wind_m0}
\end{equation}
where $f_l=0.41$ Hz and $f_u=10$ Hz are the frequency limits used in the energy calculation when wind is present.

Because the wind-generated wave field was statistically stationary, its energy over the same integration duration ($\Delta T=30$\,s) used for the focused wave groups can be evaluated from equation~\eqref{eq:eta energy} as

\begin{equation}
E_w(x)
=\rho_wg\Delta T \left\langle C_g\right\rangle_w(x)m_0^{w}(x),
\label{eq:wind_energy}
\end{equation}
where $\left\langle C_g\right\rangle_w(x)$ is calculated from the wind-only spectrum using equation~\eqref{eq:Cgs}.

Denoting by $E_m(x)$ the energy calculated from the mixed surface-elevation record using equations~\eqref{eq:eta energy} and \eqref{eq:Cgs}, the fractional contribution from the purely wind-generated wave field is

\begin{equation}
\mathcal{R}_w(x) = \frac{E_w(x)}{E_m(x)}.
\label{eq:wind_energy_fraction}
\end{equation}
The energy associated with the prescribed focused wave group is therefore

\begin{equation}
\begin{aligned}
E_p(x) &=
E_m(x)\left[1-\mathcal{R}_w(x)\right] \\
&=E_m(x)-E_w(x).
\end{aligned}
\label{eq:corrected_energy}
\end{equation}

The values of $E_p(x)$ at the inlet and outlet of the control volume, $E_p(x_I)$ and $E_p(x_O)$, are then substituted for $E(x_I)$ and $E(x_O)$, respectively, in equation~\eqref{eq:delta E 1}. The breaking-induced energy dissipation $\Delta E_{br}$ is then estimated using the refined framework described in \S\ref{subsec: Cao's method}. Any net wind work on the focused wave group within the control volume remains in the energy balance after this subtraction and therefore contributes to the wind dependence of the inferred dissipation.

\bibliographystyle{apalike}
\bibliography{Biblio}

\end{document}